\documentclass[10pt,a4paper]{article}
\usepackage[margin=2cm]{geometry}

\usepackage{graphicx}
\usepackage{amsmath}
\usepackage{amssymb}

\usepackage{dcolumn}
\usepackage{bm}

\usepackage[utf8]{inputenc}
\usepackage[T1]{fontenc}
\usepackage{mathptmx}
\usepackage{etoolbox}
\usepackage{caption}
\usepackage{subcaption}

\usepackage{xcolor}

\usepackage{authblk}

\begin{document}

\title{Phase-Amplitude Reduction and Optimal Phase Locking of Collectively Oscillating Networks}

\author[1]{Petar Mircheski\footnote{Corresponding author: mircheski.p@k.sc.e.titech.ac.jp}}
\author[2]{Jinjie Zhu}
\author[1]{Hiroya Nakao}
\affil[1]{Department of Systems and Control Engineering, Tokyo Institute of Technology, Tokyo 152-8552, Japan}
\affil[2]{State Key Laboratory of Mechanics and Control of Mechanical Structures, College of Aerospace Engineering, Nanjing University of Aeronautics and Astronautics, Nanjing 210016, China}

\maketitle
\begin{abstract}
We present a phase-amplitude reduction framework for analyzing collective oscillations in networked dynamical systems.
The framework, which builds on the phase reduction method, takes into account not only the collective dynamics on the limit cycle but also deviations from it by introducing amplitude variables and using them with the phase variable.
The framework allows us to study how networks react to applied inputs or coupling, including their synchronization and phase-locking, while capturing the deviations of the network states from the unperturbed dynamics.
Numerical simulations are used to demonstrate the effectiveness of the framework for networks composed of FitzHugh-Nagumo elements.
The resulting phase-amplitude equation can be used in deriving optimal periodic waveforms or introducing feedback control for achieving fast phase locking while stabilizing the collective oscillations.
\end{abstract}

{\bf
Networked dynamical systems can exhibit collective oscillations that are important in their functioning, such as in biological and engineered systems. We develop a phase-amplitude reduction theory for collective oscillations in networked systems by extending previous theories, which enables us to describe the collective oscillations of high-dimensional dynamical networks using low-dimensional phase-amplitude equations. As an illustration, we analyze optimal phase locking of FitzHugh-Nagumo networks and demonstrate that we can achieve faster phase locking by using the amplitude equation. In particular, we show that multi-element control outperforms single-element control in realizing stable phase locking.
}

\section{Introduction}
Synchronized collective dynamics of networked dynamical systems are commonly observed in the real world, e.g., in biological systems and engineered systems~\cite{winfree1967biological, winfree1980geometry, pikovsky2002synchronization, strogatz2012sync}. The use of dimensionality reduction methods is widespread in the analysis of complex dynamics in various systems exhibiting synchronization.
One particularly common method, known as phase reduction~\cite{winfree1980geometry, pikovsky2002synchronization, kuramoto1984chemical, hoppensteadt1997weakly, ermentrout2010mathematical, brown2004phase, nakao2016phase, monga2019phase, monga2019optimal}, is especially effective in the study of limit-cycle oscillators.
This approach simplifies the dynamics of a stable limit-cycle oscillator by projecting the state of the oscillator onto the phase direction of the unperturbed limit-cycle orbit.

When the perturbations applied to the oscillator are sufficiently weak, the phase sensitivity function~\cite{winfree1967biological,winfree1980geometry,kuramoto1984chemical} can be used to evaluate their effect on the oscillator's phase under linear approximation, resulting in a simple one-dimensional equation that describes the evolution of the phase variable.
The phase reduction technique is crucial in the examination of how oscillators interact, including mutual synchronization of coupled oscillators, phase locking of an oscillator to external forcing, and formation of rhythmic spatiotemporal patterns in chemical and biological systems~\cite{winfree1980geometry, pikovsky2002synchronization, strogatz2012sync, ermentrout2010mathematical, iatsenko2013evolution, kuramoto1984chemical}.
The phase equation presents a diverse range of uses in the control of limit-cycling systems. Its various applications include optimal phase control~\cite{moehlis2006optimal,monga2019phase,monga2019optimal}, increasing the phase-locking range~\cite{tanaka2014optimal, tanaka2015optimal, harada2010optimal, kato2021optimization}, improving the linear stability of phase locking~\cite{zlotnik2013optimal} or mutual synchronization~\cite{watanabe2019optimization}, maximizing the coherence~\cite{pikovsky2015maximizing}, and achieving phase-selective entrainment~\cite{zlotnik2016phase}.

Recently, a generalization of the conventional phase reduction method, known as phase-amplitude reduction, has been developed~\cite{wedgwood2013phase,wilson2016isostable, shirasaka2017phase, mauroy2018global, wilson2019phase, mauroy2020koopman, shirasaka2020phase, takata2021fast, wilson2020phase, kotani2020nonlinear}.
This method takes into account not only the dynamics projected on the limit cycle but also deviations from it.
Unlike phase reduction, which uses only the phase variable to describe the oscillator state relying on the assumption that the oscillator is only weakly perturbed and the oscillator state does not deviate from the unperturbed limit cycle, phase-amplitude reduction uses amplitude variables in addition to the phase variable to describe the deviations of the oscillator state from the unperturbed limit cycle.

For an exponentially stable limit-cycle oscillator, the asymptotic phase is introduced such that it increases with a constant frequency as the oscillator state evolves in the basin of the limit cycle~\cite{winfree1967biological, winfree1980geometry, pikovsky2002synchronization, kuramoto1984chemical, ermentrout2010mathematical, nakao2016phase, brown2004phase, monga2019phase}.
The amplitudes of the oscillator state can be introduced similarly to the asymptotic phase such that they decay exponentially to zero as the oscillator state converges to the unperturbed limit cycle~\cite{mauroy2013isostables,wilson2016isostable, shirasaka2017phase, mauroy2018global, wilson2019phase, mauroy2020koopman, shirasaka2020phase, takata2021fast}.
The phase and amplitudes are closely related to the Koopman eigenfunctions of the system~\cite{mauroy2013isostables, mauroy2016global, shirasaka2017phase, mauroy2018global, mauroy2020koopman}, where the Koopman eigenvalues are characterized by the natural frequency and the Floquet exponents of the limit cycle, and the level sets of the phase and amplitude functions are called isochrons and isostables, respectively.
By keeping only the slowest-decaying amplitude variable, it is possible to derive a pair of coupled equations for the phase and amplitude of a weakly perturbed limit-cycle oscillator state from its dynamical equation.
There have been recent suggestions for various uses of these phase-amplitude equations.

In this study, building upon the phase reduction method, in particular, that for networked or spatially-extended dynamical systems~\cite{kawamura2008collective,kori2009collective,nakao2014phase,nakao2018phase,nakao2021sparse,nakao2021phase}, we develop a phase-amplitude reduction framework for networks of dynamical elements that exhibit stable collective oscillations.
In addition to the phase variable characterizing the collective phase of the network, we also consider amplitude variables, which describe how the network deviates from its unperturbed collective oscillations.
We derive the coupled adjoint equations that describe how the collective phase and amplitude of the network respond to small perturbations applied to the individual elements of the network.
These phase and amplitude sensitivity functions can be used to derive a pair of coupled phase and amplitude equations describing the collective oscillations of the network under weak perturbations.

To illustrate our method, we consider collective oscillations in networks of FitzHugh-Nagumo elements with two different topologies and perform phase-amplitude reduction to study their synchronization properties~\cite{nakao2018phase,nakao2021sparse}.
The resulting phase-amplitude equation are used in deriving the optimal periodic waveforms to achieve phase-locking~\cite{moehlis2006optimal, zlotnik2013optimal, takata2021fast, kato2021optimization, harada2010optimal} while suppressing amplitude deviations from the limit cycle.
We also implement a simple feedback control of the amplitude, which was previously proposed for a single oscillator~\cite{takata2021fast}, and show that it can stabilize the collective oscillations without altering the phase dynamics.
We also demonstrate that multi-element control outperforms single-element control in both cases.

\section{Collectively oscillating networks}

We consider a network of $N$ coupled dynamical elements described by
\begin{align}
	\dot{\bm x}_{i}(t) = {\bm f}_{i}({\bm x}) + \sum_{j =1}^{N} {\bm g}_{ij}({\bm x}_{i}, {\bm x}_{j}),
	\quad (i=1, ..., N).
	\label{eq:vectorf}
\end{align}
Here, ${\bm x}_i(t) \in {\mathbb R}^{n_i}$ represents an $n_i (\geq 1)$-dimensional state of element $i$ at time $t$, overdot $\langle \dot{} \rangle$ represents time derivative,
${\bm f}_i : {\mathbb R}^{n_i} \to {\mathbb R}^{n_i}$ represents individual dynamics of element $i$, and ${\bm g}_{ij} : {\mathbb R}^{n_i} \times {\mathbb R}^{n_j} \to {\mathbb R}^{n_i}$ describes the effect of element $j$ on element $i$, respectively.
The vector fields ${\bm f}_i$ and ${\bm g}_{ij}$ are assumed to be continuously differentiable.
We assume that no element is isolated, i.e., the interaction network is a connected graph.
We denote the total dimensionality of the whole networked dynamical system by
\begin{align}
	M = \sum_{i=1}^N n_i.
\end{align}
Let ${\bm X} = (\bm{x}^{\top}_{1}, \bm{x}^{\top}_{2}, \ldots, \bm{x}^{\top}_{N})^{\top} \in {\mathbb R}^M$, where $\top$ indicates matrix transpose, collectively represents the network state as an $M$-dimensional column vector. Then Eq.~\eqref{eq:vectorf} can also be expressed collectively as a single $M$-dimensional dynamical system $\dot{\bm X}(t) = {\bm F}({\bm X})$ where ${\bm F} : {\mathbb R}^M \to {\mathbb R}^M$ is determined from ${\bm f}_i$ and ${\bm g}_{ij}$ (see Appendix A).
The dimensionality of each individual element of the network can vary and the coupling of the network can be arbitrary, as long as the network displays linearly stable limit-cycle dynamics.

We assume that the network described by Eq.~\eqref{eq:vectorf} exhibits stable collective oscillations, in which each individual element behaves periodically, i.e., $\tilde{\bm x}_i(t) = \tilde{\bm x}_i(t+T)$ for $i=1, ..., N$ with a natural period $T$ and frequency $\omega = 2\pi / T$.
Specifically, we assume that Eq.~\eqref{eq:vectorf} has an exponentially stable limit cycle solution ${\bm X}(t) = \tilde {\bm X}(t) = (\tilde{\bm x}^{\top}_1(t), \tilde{\bm x}^{\top}_2(t), ..., \tilde{\bm x}^{\top}_N(t))^{\top} \in {\mathbb R}^M$, satisfying $\tilde {\bm X}(t) = \tilde {\bm X}(t + T)$.
The linear stability of this limit-cycle solution is characterized by its Floquet exponents $\lambda_{0}, \lambda_1, \ldots, \lambda_{M-1}$.
One of the Floquet exponents is zero, $\lambda_{0} = 0$, which is associated with the tangent direction along the limit cycle of the network.
The other exponents $\lambda_1, \ldots, \lambda_{N-1}$, which are either real or complex, are associated with the amplitude directions away from the limit cycle and possess negative real parts.

We note that each element may not necessarily be oscillatory when isolated, but the network exhibits stable collective oscillations as a whole. For example, in the example of the ring network introduced later, all elements are excitatory rather than oscillatory, but the whole network exhibits collective oscillations in the form of traveling waves.

\section{FitzHugh-Nagumo networks}

\subsection{Networks of FitzHugh-Nagumo elements}

The phase-amplitude reduction framework for collectively oscillating networks that we develop in this study is general, but as concrete examples,
we consider networks of interconnected FitzHugh-Nagumo elements
with two types of topologies.
The state of the $i$-th element is given by a column vector ${\bm x}_i(t) = (u_{i}(t), v_i(t))^{\top}$ for $i=1, ..., N$, and the individual dynamics ${\bm f}_i(u_i, v_i)$ of the $i$-th element is given by a two-dimensional vector field
\begin{equation}
	\begin{pmatrix} \dot{u}_{i} \\ \dot{v}_{i}\end{pmatrix} =
	\begin{pmatrix}\delta (v_{i} + a + b u_{i}) \\ v_{i} - \frac{v^{3}_{i}}{3} - u_{i} + I_{i}\end{pmatrix},
\end{equation}
where $\delta$, $a$, and $b$ are parameters identical for all the elements, and $I_{i}$ is the excitation current being applied to the $i$-th element.
Each element of the network can be excitatory or oscillatory depending on the value of $I_{i}$.
Because the FitzHugh-Nagumo model is a neuron model, we assume that the interaction between the elements can only occur through their membrane potential. Specifically, we define the coupling function between the elements $i$ and $j$ by
\begin{align}
\bm{g}_{ij}({\bm x}_i, {\bm x}_j) = K_{ij}\begin{pmatrix}0 \\ v_{j} - v_{i}\end{pmatrix},
\end{align}
where $K_{ij}$  represents the coupling strength between the elements $i$ and $j$ ($i, j = 1, \ldots, N$).

Two cases for the coupling and excitation current have been taken into account: (i) a ring network where each element is coupled only to its adjacent two elements with periodic boundary conditions and (ii) a random network where all elements are randomly coupled.
The parameters $a=0.7$, $b=0.8$, and $\delta=0.08$ are identical while the excitation current $I_{i}$ and the coupling strengths $\{ K_{ij} \}$ differ depending on the networks.
We use the numerical algorithms described in Ref.~\cite{takata2021fast} to calculate the Floquet exponents that characterize the linear stability of the limit cycle corresponding to their collective oscillations.

\subsection{Ring network with local coupling}

\begin{figure*}[htbp]
    \includegraphics[width=\linewidth]{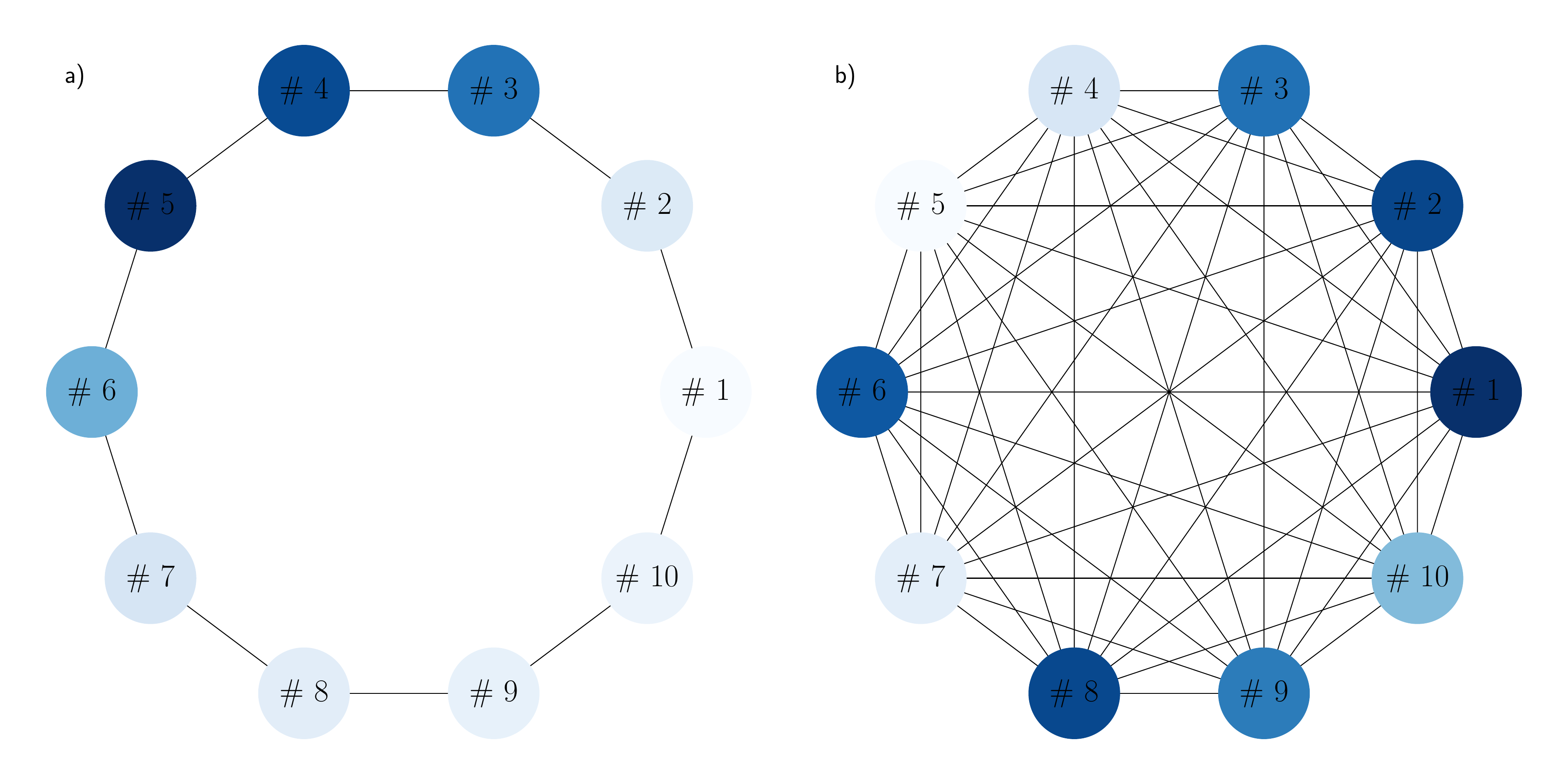}
	\caption{(a)
	  Connectivity of the FizHugh-Nagumo ring network.
	  (b) Connectivity of the FizHugh-Nagumo random network. Each vertex color represents the value of the state variable $v_i$ of the element $i$ at the collective phase $\theta = \pi$.}
	\label{fig:connectivity}
\end{figure*}

\begin{figure*}[htbp]
    \includegraphics[width=\linewidth]{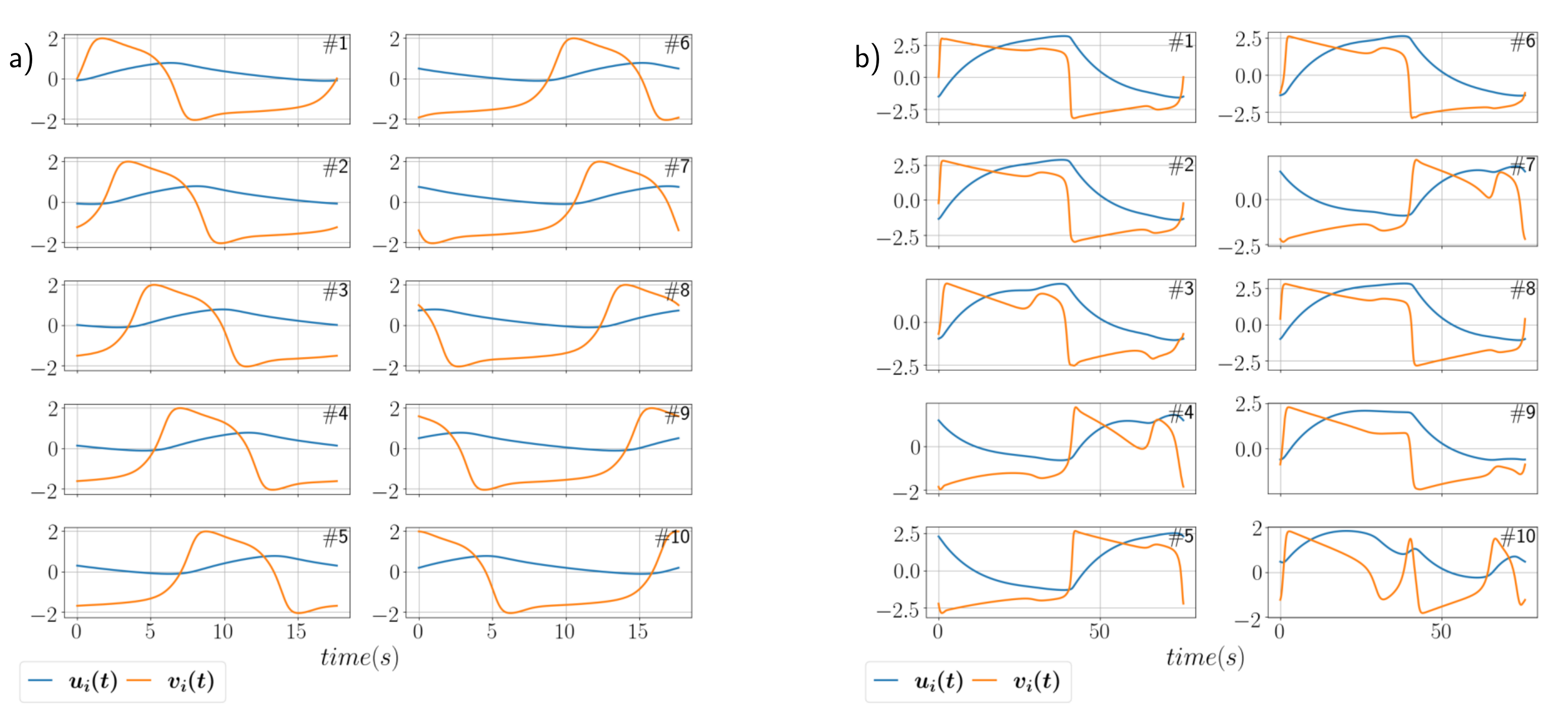}
	\caption{(a)
	  One-period time evolution of ${\bm x}_i(t) = (u_{i}(t), v_i(t))^{\top}$ for the ring network.
	  (b) One-period time evolution of ${\bm x}_i(t) = (u_{i}(t), v_i(t))^{\top}$ for the random network.}
	\label{fig:limit_cycle}
\end{figure*}

As the first example of the network, we consider a locally coupled network of $N=10$ excitatory FitzHugh-Nagumo elements, where every element of the network is coupled only to its two adjacent elements with periodic boundary conditions (Fig.~\ref{fig:connectivity}(a)).
The matrix of coupling strengths is given by $K_{i,i-1} = +0.3$, $K_{i,i+1} = -0.3$, and $K_{i,j} = 0$ otherwise, where the elements $0$ and $N+1$ are identified with the elements $N$ and $1$, respectively.
The input currents to the elements are fixed as $I_{1-10} = 0.32$ for this network, hence all elements are excitatory.
The limit-cycle solution of the ring network is a pulse that travels through the elements of the network in one period of oscillation (Fig.~\ref{fig:limit_cycle}(a)), which can be obtained by choosing appropriate initial conditions.

We have numerically evaluated the natural period of this network to be $T\approx17.66$ and accordingly, the natural frequency of the network to be $\omega \approx 0.355$. The slowest decaying Floquet exponents have been calculated as $\lambda_{1,2} = -0.022 \pm 0.171 \imath$, and the second slowest exponents as $\lambda_{3,4}=-0.067 \pm 0.1 \imath$.
It is noted that the slowest exponents are a complex-conjugate pair and therefore the amplitude exhibits oscillatory decay towards the limit cycle.

\subsection{Network with random coupling}

As another example of the network, we consider a randomly coupled network of $N=10$ excitatory and oscillatory FitzHugh-Nagumo elements (Fig.~\ref{fig:connectivity}(b)), which was analyzed in Refs.~\cite{nakao2018phase,nakao2021sparse}.
We use the matrix of random coupling strengths given in Appendix B.
The input currents are fixed as $I_{1-7} = 0.2$ (excitatory) and $I_{8-10} = 0.8$ (oscillatory).
Since the elements are randomly coupled and their parameters are different, they do not exhibit uniformly synchronized oscillations, but all the elements are synchronous in the sense that each element regularly exhibits periodic dynamics while keeping constant timing with the other elements (Fig.~\ref{fig:limit_cycle}(b)).
Such time-locked dynamics of neurons are called 'polychronization' in Ref.~\cite{izhikevich2006polychronization}.
This is the limit-cycle solution corresponding to the collective oscillations of this random network.

We have numerically evaluated the natural period of this network as $T \approx 75.711$ and accordingly the natural frequency  as $\omega \approx 0.083$.
The slowest-decaying Floquet exponent has been calculated as $\lambda_{1} = -0.058$, and the second-slowest exponents as $\lambda_{2,3} = -0.088 \pm 0.0027 \imath$.
The slowest exponent is real and the amplitude exhibits a simple exponential decay in this case.

\section{Phase-amplitude reduction}

We now formulate a phase-amplitude reduction method for networks of dynamical elements exhibiting collective oscillations, extending the phase reduction method for networks~\cite{nakao2018phase} and the phase-amplitude reduction method for ordinary limit-cycle oscillators~\cite{takata2021fast} developed previously.
When the network is composed of many elements, the dimensionality $M$ of the network state is large. However, when the network exhibits stable collective oscillations as assumed, we can approximately describe the network dynamics under weak perturbations by using only the phase and a few amplitudes. This reduces the dimensionality of the dynamics and facilitates analysis and control of the network.

We consider a stable limit-cycle solution of the network given by Eq.~\eqref{eq:vectorf}.
In the basin $B \subset {\mathbb R}^M$ of the limit cycle $\tilde{\bm X}(t) = (\tilde{\bm x}^{\top}_{1}(t), \tilde{\bm x}^{\top}_{2}(t), \ldots, \tilde{\bm x}^{\top}_{N}(t))^{\top}$, we can define an asymptotic phase function $\Theta: B \rightarrow \left[0, 2\pi \right)$ and amplitude functions $R_{m}: B \rightarrow \mathbb{C}$ ($m=1, ..., M-1$), which map a network state
${\bm X} = (\bm{x}_{1}, \bm{x}_{2}, \ldots, \bm{x}_{N}) \in B$ to a phase and to (real or complex) amplitudes, respectively.

We assume that the phase function $\Theta({\bm X}) = \Theta({\bm x}_{1}, {\bm x}_2, \ldots, {\bm x}_{n})$ constantly increases with the frequency $\omega$ of the collective oscillations as given by the following equations:
\begin{align}
	\label{eq:phase_equation}
	 & \dot{\Theta}(\bm{x}_{1}, \bm{x}_{2}, \ldots, \bm{x}_{N}) = \sum_{i=1}^N \frac{\partial \Theta}{\partial {\bm x}_i} \cdot \dot{\bm x}_i = \sum_{i=1}^{N}\frac{\partial \Theta}{\partial {\bm x}_{i}} \cdot \left( \bm{f}_{i}({\bm x}_i) + \sum_{j =1}^{N} {\bm g}_{ij}({\bm x}_{j}, {\bm x}_{i}) \right) = \omega,
\end{align}
where $\partial/\partial {\bm x}_i$ represents the gradient with respect to ${\bm x}_i$, $\langle \cdot \rangle$ represents the ordinary scalar product of two vectors, i.e., ${\bm a} \cdot {\bm b} = \sum_{i=1}^N a_i b_i$, and we used the chain rule of differentiation and Eq.~\eqref{eq:vectorf}. We choose one point $\tilde{\bm X}^O = ({\bm x}_{1}^O, {\bm x}_2^O, \ldots, {\bm x}_{n}^O)$ on the limit cycle as the phase origin, i.e., $\Theta({\bm x}_{1}^O, {\bm x}_2^O, \ldots, {\bm x}_{n}^O) = 0$.
In what follows, we will also use the phase $\theta = \omega t \in [0, 2\pi)$ to represent a point on the limit cycle as $\tilde{\bm X}(\theta) = (\tilde{\bm x}_1(\theta), \ldots, \tilde{\bm x}_N(\theta))^{\top}$ in place of the time $t \in [0, T)$.

Similarly, we introduce the amplitude functions $R_m({\bm X}) = R_m({\bm x}_{1}, {\bm x}_2, \ldots, {\bm x}_{n})$ satisfying
\begin{align}
	\label{eq:amp_equation}
	 & \dot{R}_{m}(\bm{x}_{1}, \bm{x}_{2}, \ldots, \bm{x}_{N}) = \sum_{i=1}^N \frac{\partial R_{m}}{\partial {\bm x}_i} \cdot \dot{\bm x}_i = \sum_{i=1}^{N}\frac{\partial R_{m}}{\partial {\bm x}_{i}} \cdot \left({\bm f}_{i}({\bm x}) + \sum_{j =1}^{N} {\bm g}_{ij}({\bm x}_{j}, {\bm x}_{i}) \right) = \lambda_{m}R_{m},
\end{align}
for $m=1, ..., M-1$, where $\lambda_1, \ldots, \lambda_{M-1}$ are the  Floquet exponents of the limit cycle.
Namely, the amplitude $R_m$ characterizing the deviation from the limit cycle obeys a linear equation with the decay rate $\lambda_m$ and, when the network state is on the limit cycle, the amplitude vanishes, i.e., $R_m(\tilde{\bm x}_1, ..., \tilde{\bm x}_N) = 0$.

The phase function and amplitude functions are closely related to the Koopman eigenfunctions of the system~\cite{mauroy2020koopman,mauroy2013isostables,shirasaka2020phase};
the complex exponential of the phase function, $e^{i \Theta}$, and the amplitude functions, $R_1, ..., R_m$, are the Koopman eigenfunctions associated with the eigenvalues $i \omega$ and $\lambda_1, ..., \lambda_{M-1}$ of the limit cycle, respectively.

We now assume that each element of the network is driven by a weak external input ${\bm p}_i(t) \in {\mathbb R}^{n_i}$.
The perturbed network obeys
\begin{equation}
	\dot{\bm x}_{i}(t) = {\bm f}_{i}({\bm x}_{i}) + \sum_{j=1}^{N} {\bm g}_{ij}({\bm x}_{j}, {\bm x}_{i}) + {\bm p}_{i}(t),
	\quad (i=1, ..., N).
	\label{eq:perturbated}
\end{equation}
We assume that ${\bm p}_i(t)$ is sufficiently weak such that the collective oscillations of the network persist and the network state does not deviate from the unperturbed limit cycle too largely.

We define the phase and amplitude variables of this network by $\theta(t) = \Theta({\bm X}(t)) =\Theta({\bm x}_1(t), ..., {\bm x}_N(t))$ and $ r_{m}(t) = R_{m}(\bm X(t)) = R_{m}({\bm x}_1(t), ..., {\bm x}_N(t))$ for $m=1, ..., M-1$.
By using the chain rule of differentiation and Eq.~\eqref{eq:perturbated}, we can derive the following equations for the phase and amplitude variables:
\begin{align}
	 & \dot{\theta}(t) = \omega + \sum_{i=1}^{N} \frac{\partial \Theta}{\partial {\bm x}_{i}} \cdot {\bm p}_{i}(t),
	 \\
	 & \dot{r}_{m}(t) = \lambda_{m} r_{m}(t) + \sum_{i=1}^{N}\frac{\partial R_{m}}{\partial {\bm x}_{i}} \cdot {\bm p}_{i}(t),
\end{align}
for $m=1, ..., M-1$. These equations are not yet closed in $\theta$ and $r_1, ..., r_{M-1}$ because of the gradient terms that explicitly depend on ${\bm X} = ({\bm x}_1, ..., {\bm x}_N)$.

To obtain the phase-amplitude equations in a closed form, we assume that the weak inputs ${\bm p}_1, ..., {\bm p}_N$ are of order $O(\varepsilon)$ ($0 \leq \varepsilon \ll 1$) and the deviation of the state of each element ${\bm x}_i$ from the unperturbed state $\tilde{\bm x}_{i}$ is also of order $O(\varepsilon)$. We can then approximately evaluate the gradient terms on the limit cycle at $\tilde{\bm X}(\theta) =(\tilde{\bm x}_1(\theta), \ldots, \tilde{\bm x}_N(\theta))$ as
$\partial \Theta / \partial {\bm x}_i = {\bm z}_i(\theta) + O(\varepsilon)$ and $\partial R_m / \partial {\bm x}_i = {\bm I}_{m,i}(\theta) + O(\varepsilon)$,
where we defined
\begin{align}
&{\bm z}_i(\theta) = \left. \frac{\partial \Theta({\bm x}_1, ..., {\bm x}_N)}{\partial {\bm x}_{i}} \right|_{(\tilde{\bm x}_1(\theta), ..., \tilde{\bm x}_N(\theta))} \in {\mathbb R}^{n_i},
\end{align}
and
\begin{align}
&{\bm I}_{m,i}(\theta) = \left. \frac{\partial R_m({\bm x}_1, ..., {\bm x}_N)}{\partial {\bm x}_{i}} \right|_{(\tilde{\bm x}_1(\theta), ..., \tilde{\bm x}_N(\theta))} \in {\mathbb R}^{n_i}.
\end{align}
We call $\bm{z}_{i}$ and ${\bm I}_{m,i}$  the phase (isochron) sensitivity function (PSF) and the isostable (amplitude) sensitivity functions (ISFs), respectively.
We note that they are $2\pi$-periodic functions of $\theta$ by definition.
Neglecting the terms of $O(\varepsilon^2)$, we obtain a set of approximate phase-amplitude equations:
\begin{align}
	 & \dot{\theta}(t) = \omega +\sum_{i=1}^{N} {\bm z}_{i}(\theta) \cdot {\bm p}_{i}(t),                                        \cr
	 & \dot{r}_{m}(t) = \lambda_{m} r_{m}(t) + \sum_{i=1}^{N} {\bm I}_{m, i}(\theta) \cdot {\bm p}_{i}(t),
	\label{eq:psf_isf}
\end{align}
which are closed in the phase $\theta$ and amplitudes $r_1, ..., r_{M-1}$ and correct up to $O(\varepsilon)$, i.e., to the first order in the external input.
The PSF and ISFs characterize the linear response properties of the phase and amplitudes of the collective oscillation to small inputs given to individual elements. Although it is difficult to obtain $\Theta$ and $R_m$ explicitly, the PSF ${\bm z}_i$ and ISFs $\{ {\bm I}_{m,i} \}$ can be numerically evaluated by solving the adjoint equations as explained in the next section.

We have so far retained all $M$ degrees of freedom. For typical limit-cycling networks, it is often the case that many Floquet exponents have small real parts, namely, the corresponding modes quickly decay to zero and only $\lambda_0 = 0$ associated with the phase variable and a few Floquet exponents associated with slowly-decaying amplitude variables are practically important. In such cases, by keeping only the $n$ ($< M-1$) slowest decaying modes of the network, we can effectively reduce the dimensionality of the network.
In particular, if we retain only the slowest-decaying amplitude with the Floquet exponent $\lambda_1$, we obtain a pair of phase-amplitude equations:
\begin{align}
	 & \dot{\theta}(t) = \omega +\sum_{i=1}^{N} {\bm z}_{i}(\theta) \cdot {\bm p}_{i}(t),
	 \cr
	 & \dot{r}_{1}(t) = \lambda_{1} r_{1}(t) + \sum_{i=1}^{N} {\bm I}_{1, i}(\theta) \cdot {\bm p}_{i}(t).
	\label{eq:psf_isf_2}
\end{align}
We note that $r_1$ is a complex variable when $\lambda_1$ is complex.
Thus, we have reduced the dimensionality of the network from $M$ to $2$ (when $\lambda_1$ is real) or $3$ (when $\lambda_1$ is complex).

\section{Adjoint equations for the sensitivity functions}

The phase-amplitude equations~\eqref{eq:psf_isf} are characterized by the natural frequency $\omega$, Floquet exponents $\{ \lambda_m \}$, PSFs $\{ {\bm z}_i(\theta) \}$, and ISFs $\{ {\bm I}_{m,i}(\theta) \}$ for $m=1, ..., M-1$ and $i=1, ..., N$.
In practice, it is difficult, if not impossible, to obtain the phase function $\Theta$ and amplitude function $R_{m}$ explicitly. However, their gradients evaluated on the limit cycle, namely, the PSF and ISFs, can be obtained by solving the adjoint equations as explained below.
The adjoint equation for the PSF has been derived in Ref.~\cite{nakao2018phase} by using the method in Ref.~\cite{brown2004phase}, which can also be considered a discrete version of the adjoint equation for reaction-diffusion systems in Ref.~\cite{nakao2014phase,nakao2021phase}. We here use the same idea to derive the adjoint equations for PSF and also ISFs.

We consider a reference network state starting from the phase origin $\tilde{\bm X}(t=0) = \tilde{\bm X}^O$ at $t=0$, whose phase $\theta = \Theta(\tilde{\bm X}^O)=0$.
We also consider a network state ${\bm X}(t) = ({\bm x}^{\top}_1(t), ..., {\bm x}^{\top}_N(t))^{\top}$ near this reference state $\tilde{\bm X}(t) = (\tilde{\bm x}^{\top}_{1}(t), ..., \tilde{\bm x}^{\top}_N(t))^{\top}$ at $t$, represented by
\begin{align}
{\bm x}_i(t) = \tilde{\bm x}_i(t) + {\bm y}_i(t)
\end{align}
for $i=1, ..., N$, where ${\bm y}_i \in {\mathbb R}^{n_i}$ is a small variation. Plugging into Eq.~\eqref{eq:vectorf}, the linearized equation for ${\bm y}_i$ is obtained as
\begin{align}
	&\dot{\bm y}_i(t) = J_i(t) {\bm y}_i(t) + \sum_{j=1}^N J_{ij}(t) {\bm y}_j(t),
	\label{eq:linvar}
\end{align}
where $T$-periodic Jacobian matrices $J_i(t) \in {\mathbb R}^{n_i \times n_i}$ and $J_{ij}(t) \in {\mathbb R}^{n_i \times n_j}$ are given by
\begin{align}
 &J_{i}(t) = \left[ D_0{\bm f}_i(\tilde{\bm x}_i(t)) + \sum_{j =1}^{N} D_{1} {\bm g}_{ij}(\tilde{\bm x}_i(t), \tilde{\bm x}_j(t)) \right],
 \cr
 &J_{ij}(t) = D_{2} {\bm g}_{ij}(\tilde{\bm x}_i(t), \tilde{\bm x}_j(t)).
\end{align}
Here, $D_0, D_1, D_2$ are gradient operators defined as
\begin{align}
	&D_0 {\bm f}_i({\bm x}_i) = \left. \frac{\partial {\bm f}_i({\bm x})}{\partial {\bm x}} \right|_{{\bm x} = {\bm x}_i},
\end{align}
and
\begin{align}
	&D_1 {\bm g}_{ij}({\bm x}_i, {\bm y}_j)= \left. \frac{\partial {\bm g}_{ij}({\bm x}, {\bm y})}{\partial {\bm x}} \right|_{({\bm x}, {\bm y}) = ({\bm x}_i, {\bm y}_j)},
	\cr
	&D_2 {\bm g}_{ij}({\bm x}_i, {\bm y}_j) = \left. \frac{\partial {\bm g}_{ij}({\bm x}, {\bm y})}{\partial {\bm y}} \right|_{({\bm x}, {\bm y}) = ({\bm x}_i, {\bm y}_j)}.
\end{align}

We consider the asymptotic phase and amplitudes of the network state $( {\bm x}_1, ..., {\bm x}_N )$. The phase and amplitude functions can be expanded to the first order in ${\bm y}_i$ as
\begin{align}
&
\Theta({\bm x}_1, ..., {\bm x}_N)
= \Theta( \tilde{\bm x}_1 + {\bm y}_1, ..., \tilde{\bm x}_N + {\bm y}_N)
\approx \Theta(\tilde{\bm x}_1, ..., \tilde{\bm x}_N) + \sum_{i=1}^N {\bm z}_i \cdot {\bm y}_i,
\cr
&R_m({\bm x}_1, ..., {\bm x}_N)
= R_m( \tilde{\bm x}_1 + {\bm y}_1, ..., \tilde{\bm x}_N + {\bm y}_N)
\approx
\sum_{i=1}^N {\bm I}_{m,i} \cdot {\bm y}_i,
\end{align}
where
we used that ${\bm z}_i$ and ${\bm I}_{m,i}$ are the gradients of
$\Theta$ and $R_m$ evaluated on the limit cycle at $(\tilde{\bm x}_1, ..., \tilde{\bm x}_N)$ and that $R_m(\tilde{\bm x}_1, ..., \tilde{\bm x}_N) = 0$ on the limit cycle.

First, differentiating $\Theta$ by time $t$, we obtain
\begin{align}
&\frac{d}{dt} \Theta({\bm x}_1, ..., {\bm x}_N)
\approx \frac{d}{dt}
\Theta(\tilde{\bm x}_1, ..., \tilde{\bm x}_N) + \sum_{i=1}^N \left( \frac{d{\bm z}_i}{dt} \cdot {\bm y}_i + {\bm z}_i \cdot \frac{d{\bm y}_i}{dt} \right)
\cr
&=
\omega + \sum_{i=1}^N  \frac{d{\bm z}_i}{dt} \cdot {\bm y}_i + \sum_{i=1}^N {\bm z}_i \cdot \left( J_i {\bm y}_i + \sum_{j=1}^N J_{ij} {\bm y}_j \right)
\cr
&=
\omega + \sum_{i=1}^N \left( \frac{d{\bm z}_i}{dt}  + J_i^{\top} {\bm z}_i  + \sum_{j=1}^N J_{ji}^{\top} {\bm z}_j \right) \cdot {\bm y}_i,
\end{align}
where we used $\dot\Theta = \omega$ and exchanged $i$ and $j$ in the double sum. Now, since $\dot\Theta = \omega$ should also be satisfied for arbitrary small ${\bm y}_i$ obeying Eq.~\eqref{eq:linvar}, the PSFs should satisfy the following adjoint equation for $i=1, ..., N$:
\begin{align}
&\frac{d{\bm z}_i(t)}{dt}  + J_i(t)^{\top} {\bm z}_i(t)  + \sum_{j=1}^N J_{ji}(t)^{\top} {\bm z}_j(t)= 0.
\end{align}

Similarly, differentiating the amplitude $R_m$ by time $t$, we obtain
\begin{align}
&\frac{d}{dt} R_m({\bm x}_1, ..., {\bm x}_N)
\approx
\sum_{i=1}^N \left( \frac{d{\bm I}_{m,i}}{dt} \cdot {\bm y}_i + {\bm I}_{m,i} \cdot \frac{d{\bm y}_i}{dt} \right)
\cr
&=
\sum_{i=1}^N  \frac{d{\bm I}_{m,i}}{dt} \cdot {\bm y}_i + \sum_{i=1}^N {\bm I}_{m,i} \cdot \left( J_i {\bm y}_i + \sum_{j=1}^N J_{ij} {\bm y}_j \right)
\cr
&=
\sum_{i=1}^N \left( \frac{d{\bm I}_{m,i}}{dt}  + J_i^{\top} {\bm I}_{m,i} + \sum_{j=1}^N J_{ji}^{\top} {\bm I}_{m,j} \right) \cdot {\bm y}_i.
\end{align}
Now, since $\dot R_m = \lambda_m R_m$ should be satisfied,
\begin{align}
\sum_{i=1}^N \left( \frac{d{\bm I}_{m,i}}{dt}  + J_i(t)^{\top} {\bm I}_{m,i} + \sum_{j=1}^N J_{ji}(t)^{\top} {\bm I}_{m,j} \right) \cdot {\bm y}_i
=
\lambda_m \sum_{i=1}^N {\bm I}_{m,i} \cdot {\bm y}_i.
\end{align}
should hold for arbitrary ${\bm y}_i$. Thus, the ISFs of the elements $i=1, ..., N$ should satisfy the following adjoint equations for $m=1, ..., M-1$:
\begin{align}
&\frac{d{\bm I}_{m,i}(t)}{dt}  + J_i(t)^{\top} {\bm I}_{m,i}(t) + \sum_{j=1}^N J_{ji}(t)^{\top} {\bm I}_{m,j}(t) = \lambda_m {\bm I}_{m, i}(t).
\end{align}
If the above adjoint equations are satisfied, $\dot\Theta = \omega$ and $\dot R_m = \lambda_m R_m$ are satisfied for arbitrary ${\bm y}_i$ up to the first order.

We have so far represented the quantities as functions of time $t$. When expressed with the phase $\theta = \omega t$ of the reference orbit, the adjoint equations are expressed as
\begin{align}
&\omega \frac{d{\bm z}_i(\theta)}{d\theta}  + J_i(\theta)^{\top} {\bm z}_i(\theta)  + \sum_{j=1}^N J_{ji}(\theta)^{\top} {\bm z}_j(\theta)= 0,
\end{align}
\begin{align}
&\omega \frac{d{\bm I}_{m,i}(\theta)}{d\theta}  + J_i(\theta)^{\top} {\bm I}_{m,i}(\theta) + \sum_{j=1}^N J_{ji}(\theta)^{\top} {\bm I}_{m,j}(\theta) = \lambda_m {\bm I}_{m, i}(\theta),
\end{align}
where the quantities are now regarded as functions of the phase $\theta$. The PSF and ISFs are found as $2\pi$-periodic solutions to these adjoint equations.
We note that, since we only consider a single amplitude, we need to solve the adjoint equations only for ${\bm z}_i(\theta)$ and ${\bm I}_{1, i}(\theta)$ with $i=1, ..., N$.

As shown in the Appendix, we can also derive the adjoint equations by considering the whole network as a large single limit-cycle oscillator, which gives the adjoint equations for the collectively expressed PSF ${\bm Z} = ({\bm z}_1, \ldots, {\bm z}_N)$ and ISFs ${\bm H}_m = ({\bm I}_{m, 1}, \ldots, {\bm I}_{m, N})$. The merit of writing the adjoint equations in the above element-wise form is that we can avoid the evaluation of $M \times M$ Jacobian matrices for the collective variables, which can be considerably large when the number $N$ of the elements is large.

\section{Optimal phase locking with amplitude stabilization}

\subsection{Phase-locking by external forcing}

One of the possible applications of the phase-amplitude equations is the phase-locking of the network by a periodic external input and its optimization to improve the stability and convergence to a phase-locked state.
We analyze the case where some of the elements in the network are driven periodically.
We denote by $S$ the set of elements receiving inputs and assume that each element $i \in S$ is subjected to a weak external periodic input ${\bm p}_i \in {\mathbb R}^{n_i}$ of $O(\epsilon)$.
The period of the input is $T_{e}$ and frequency $\Omega = {2\pi} / {T_{e}}$  , i.e.,
\begin{equation}
	{\bm p}_i(\Omega t) = {\bm p}_i(\Omega (t + T_{e})).
\end{equation}
We assume that $T_{e}$ is  close to the natural frequency $T$ of the network such that $\Omega - \omega = O(\varepsilon)$.

The lowest-order approximate phase-amplitude equations~\eqref{eq:psf_isf_2} for this system are given by
\begin{align}
	 & \dot{\theta}(t) = \omega +\sum_{i \in S} {\bm z}_{i}(\theta) \cdot {\bm p}_{i}(\Omega t),
	 \cr
	 & \dot{r}_{1}(t) = \lambda_{1} r_{1}(t) + \sum_{i \in S} {\bm I}_{i}(\theta) \cdot {\bm p}_{i}(\Omega t).
\end{align}
Here and in what follows, we write the ISFs $\{ {\bm I}_{1, i} \}$ of the slowest-decaying amplitude $R_1$ as $\{ {\bm I}_{i} \}$ for simplicity.
Following the standard phase-locking analysis \cite{kuramoto1984chemical, hoppensteadt1997weakly, nakao2016phase}, we introduce a slow relative phase $\phi = \theta - \Omega t$,  which satisfies
\begin{align}
	 & \dot{\phi} = \Delta + \sum_{i \in S} {\bm z}_{i}(\phi + \Omega t ) \cdot {\bm p}_{i}(\Omega t),
\end{align}
where $\Delta = \omega - \Omega$ is the frequency mismatch between the network and the periodic input,
and to remove the explicit time dependency of the equation, using the assumption that the frequency mismatch between the network and the periodic input is of $O(\varepsilon)$, we average the right-hand side of the phase equation over one period of oscillation \cite{kuramoto1984chemical} and obtain
\begin{align}
	 & \dot{\phi} = \Delta + \Gamma(\phi),
	 \cr
	 & \Gamma(\phi) = \sum_{i \in S} \Gamma_i(\phi),
	 \cr
	 & \Gamma_i(\phi) = \frac{1}{T_{e}}  \int_{0}^{T_{e}}{\bm z}_{i}(\phi + \Omega s) \cdot {\bm p}_{i}(\Omega s) ds,
	\label{eq:phase_diff}
\end{align}
where $\Gamma_{i}(\phi)$  represents the phase coupling function of the element $i \in S$ to the periodic external input.
The phase-locking point $\phi^{*}$ is characterized by one of the stable fixed points of
 Eq.~\eqref{eq:phase_diff} and the linear stability of $\phi^{*}$ is determined by the sum of the negative slopes of $\Gamma_{i}(\phi)$ evaluated at $\phi = \phi^{*}$, i.e.,
\begin{equation}
	- \Gamma^{'}(\phi^{*}) = -\sum_{i \in S} \Gamma^{'}_{i}(\phi^{*}).
\end{equation}
The convergence speed of the relative phase $\phi$ to the phase-locking point $\phi^*$ in the linear regime is characterized by $- 1 / \Gamma'(\phi^*)$.

\subsection{Optimization of external forcing with amplitude suppression}

As discussed in Ref.~\cite{zlotnik2013optimal} for a single oscillator, using the phase coupling function $\Gamma$, we can formulate an optimization problem for the periodic input ${\bm p}_i$ for $i \in S$ that (i) guarantees the existence of a phase-locking point $\phi^{*}$ to which the relative phase eventually converges, given by $\Delta + \Gamma(\phi^{*}) = 0$, and (ii) maximize the linear stability
under a constraint on the average power of the external input over one period,
\begin{align}
\sum_{i \in S} \left[ \| {\bm p}_{i}(t) \|^2 \right]_t
= P,
\end{align}
where $P > 0$ is the power, $\| {\bm p}_i(t) \| = \sqrt{ {\bm p}_{i}(t) \cdot {\bm p}_{i}(t)}$, and $[ f(t) ]_{t} = (1/T_e) \int_0^{T_e} f(s) ds$ denotes an average of the function $f(t)$ over one period $T_e$.
This problem can be formulated as
\begin{align}
	\text{max:} \quad & -\Gamma^{'}(\phi^{*}), \nonumber  \\
	\text{s.t:} \quad & \sum_{i \in S} \left[ \| {\bm p}_{i}(t) \|^2 \right]_t = P, \quad  \Delta + \Gamma(\phi^{*}) = 0.
	\label{eq:phase_optim}
\end{align}
The above optimization problem for a single oscillator can be solved as follows~\cite{zlotnik2013optimal}.
Noting that $\left[ {\bm z}_i'(\phi^* + \Omega t) \cdot {\bm z}_i(\phi^* + \Omega t) \right]_t = 0$ because ${\bm z}_i$ is a $2\pi$-periodic function, we obtain
\begin{align}
  &{\bm p}_i(\Omega t) = -\frac{1}{2\nu} {\bm z}_i^{'}(\phi^{*} + \Omega t) + \frac{\mu}{2\nu} {\bm z}_i(\phi^{*} + \Omega t) \quad (i \in S),
\end{align}
and
\begin{align}
  &\mu = \frac{2 \nu \Delta}{ \sum_{i \in S} [ \| {\bm z}_i(t) \|^2 ]_t }, \quad
  \nu = \frac{1}{2} \sqrt{\frac{\sum_{i \in S} [ \| {\bm z}_i' \|^2 ]_t}{P - \frac{\Delta^{2}}{\sum_{i \in S} [\| {\bm z}_i \|^2 ]_t}}},
\end{align}
by the method of Lagrange multipliers, which maximizes the linear stability $-\Gamma'(\phi^*)$ under the given constraint on the power $P$.

When the periodic inputs are sufficiently weak, the above optimal inputs realize faster synchronization than, for example, simple sinusoidal inputs. However, when the inputs are not sufficiently weak, the network state can deviate from the unperturbed limit cycle and the phase-only reduction method may fail to describe the collective oscillations of the network.
In Ref.~\cite{takata2021fast}, two methods, the amplitude penalty method and amplitude feedback method, have been proposed for a single oscillator by using the reduced amplitude equation. In this study, we extend these methods to the collective oscillations of the network and examine their efficiency.

In the amplitude penalty method, by including the ISFs of the system in the cost function, we penalize the possible deviations of the   network state from the limit cycle.
This optimization problem can be posed as
\begin{align}
	\text{max:} \quad & -\Gamma^{'}(\phi^{*}) - \sum_{i \in S} W_{i} \left[ \left| {\bm I}_{i}(\phi^*+\Omega t) \cdot {\bm p}_i(\Omega t) \right|^2 \right]_t, \nonumber \\
	\text{s.t:} \quad & \sum_{i \in S} \left[ \| {\bm p}_{i}(t) \|^2 \right]_t = P, \quad \Delta + \Gamma(\phi^{*}) = 0, \label{eq:amp_suppression}
\end{align}
where the second term  in the objective function is the amplitude penalty  that comes from the amplitude equation and $W_{i} \geq 0$ is the weight of the penalty to be included in the optimization.
From the extremum condition, we can obtain the optimal inputs $\{ {\bm p}_i \}$ for $i \in S$ as
\begin{align}
&{\bm p}_i(\Omega t)= \frac{1}{2} \left( \nu {\bm e}_i + W_i\ \mbox{Re}\ {\bm I}_{i}(\phi^* + \Omega t) {\bm I}^{\dag}_{i}(\phi^* + \Omega t) \right)^{-1} \left\{ - {\bm z}_i'(\phi^* + \Omega t) + \mu {\bm z}_i(\phi^* + \Omega t) \right\},
\end{align}
where $\dag$ denotes the Hermitian conjugate (conjugate transpose) of the matrix and the Lagrange multipliers $\mu$ and $\nu$ should satisfy
\begin{align}
&\Delta + \sum_{i \in S} \left[ {\bm z}_i \cdot \frac{1}{2} \left( \nu {\bm e}_i + W_i\ \mbox{Re}\ {\bm I}_{i} {\bm I}^{\dag}_{i} \right)^{-1} \left( - {\bm z}_i' + \mu {\bm z}_i \right) \right]_t = 0,
\end{align}
and
\begin{align}
&\mu = \frac{ \sum_{i \in S}  \left[ {\bm z}_i \cdot \left( \nu {\bm e}_i + W_i\ \mbox{Re}\ {\bm I}_{i} {\bm I}^{\dag}_{i} \right)^{-1} {\bm z}_i' \right]_t - 2 \Delta }{ \sum_{i \in S}  \left[  {\bm z}_i \cdot \left( \nu {\bm e}_i + W_i\ \mbox{Re}\  {\bm I}_{i} {\bm I}^{\dag}_{i} \right)^{-1} {\bm z}_i  \right]_t },
\end{align}
where ${\bm e}_i$ is an $n_i \times n_i$ identity matrix and $\mbox{Re}$ applies to each individual component of the matrix.

It is expected that the above result gives compromised inputs that maximize the linear stability of the phase-locking point as much as possible while avoiding large amplitude deviations of the network state from the unperturbed limit cycle.
Although it is difficult to find $\nu$ and $\mu$ analytically, we can numerically find the correct value for $\nu$, which then determines $\mu$ and $\{ {\bm p}_i \}$ for $i \in S$ such that the power constraint given in Eq.~\eqref{eq:amp_suppression} is satisfied.

Finally, in the amplitude feedback method, the optimization problem~\eqref{eq:phase_optim} remains the same as the phase-only case, but a feedback input
\begin{equation}
	{\bm p}_i^{fb}(t) = {\bm p}_i^{ls}(t) - \alpha {\bm y}_i(t)
	\quad
	(i \in S)
\end{equation}
is added to the optimized periodic input, where $\alpha \geq 0$ is the feedback gain and ${\bm y}_i(t) = {\bm x}_i(t) - \tilde{\bm x}_i(\theta(t))$ with $\theta(t) = \Theta({\bm x}_1(t), \ldots, {\bm x}_N(t))$ is the deviation of the network state from the state on the unperturbed limit cycle with the same asymptotic phase $\theta(t)$.
As the network states $({\bm x}_1, ..., {\bm x}_N)$ and $(\tilde{\bm x}_1(\theta), ..., \tilde{\bm x}_N(\theta))$ have the same asymptotic phase, the inner product of the PSF $\left( {\bm z}_1, \ldots, {\bm z}_N \right)$ and the feedback signal $\left( {\bm y}_1, \ldots, {\bm y}_N \right)$ vanishes and the addition of the feedback does not affect the phase coupling function at the lowest order.
Using this method, we can apply stronger periodic inputs to the elements for phase locking while making sure that the amplitude deviations of the network state from the limit cycle are suppressed.

We note that the amplitude penalty method gives a feedforward control, which does not require measurement of the network state, while the amplitude feedback method requires continuous measurement and evaluation of the asymptotic phase of the network state. Therefore, the amplitude penalty method requires much smaller cost if it works successfully.

\section{Numerical results}

\subsection{Phase-locking of a ring network}

\begin{figure*}[htbp]
    \includegraphics[width=\linewidth]{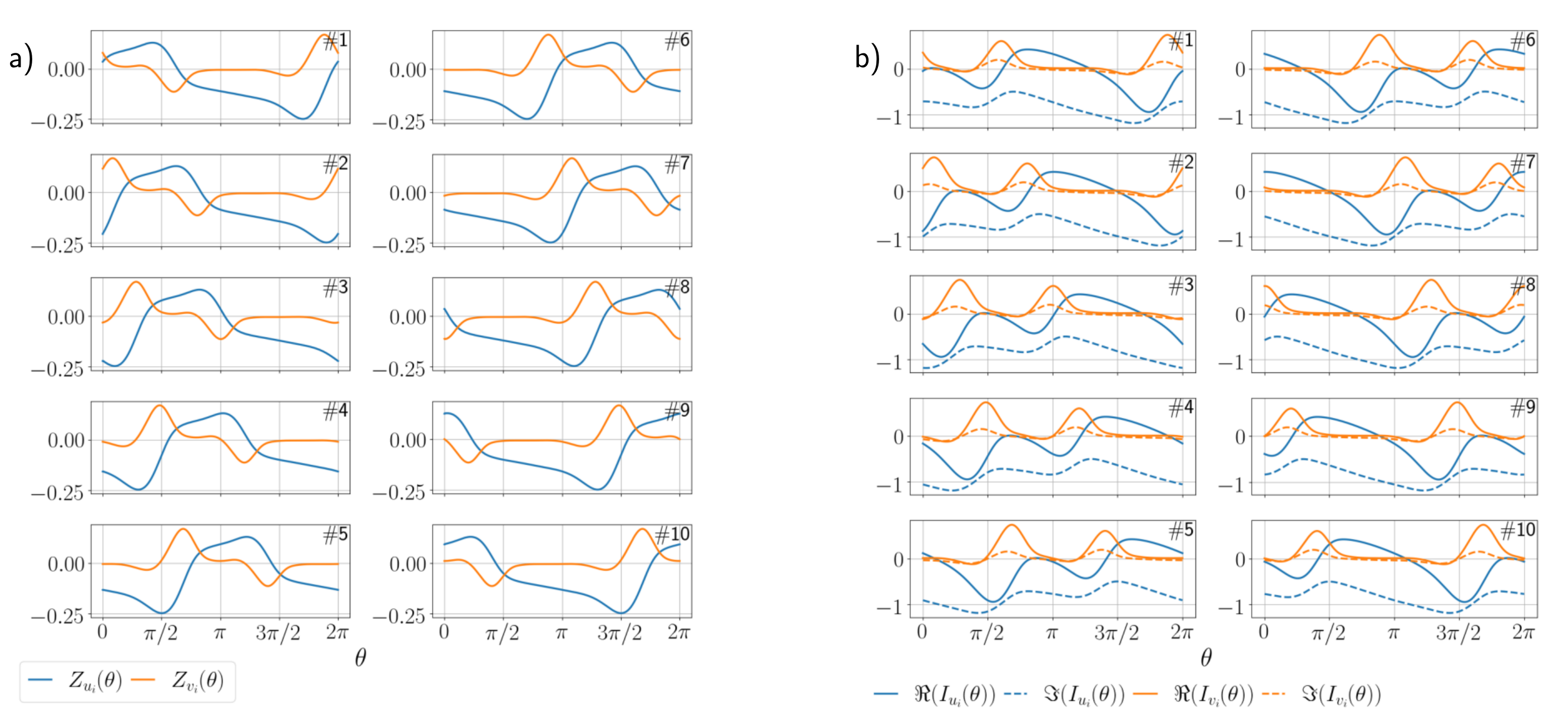}
	\caption{Phase and amplitude sensitivity functions of the ring network.
	  (a) PSFs ${\bm z}_{i}(\theta) = (z_{i}^{u}, z_{i}^{v})$.
	  (b) ISFs ${\bm I}_{i}(\theta) = ({\bm I}_{i, u}, {\bm I}_{i, v})$. Note that the amplitude sensitivity functions for the ring network are complex.}
	\label{fig:ring_psf_isf}
\end{figure*}

Numerically obtained PSFs $\{ {\bm z}_i\} $ and ISFs $\{ {\bm I}_i \}$ of the traveling-pulse solution of the ring network are shown in Figs.~\ref{fig:ring_psf_isf}(a) and (b), respectively.
Because the slowest-decaying eigenvalue is complex, the ISFs have both real and imaginary components as shown in Fig.~\ref{fig:ring_psf_isf}(b).
In this network, reflecting the translational (shift) invariance of the traveling-pulse solution on the ring, the PSFs and ISFs are also translationally invariant.
Therefore, the responses of all elements are equivalent, but we can utilize their phase differences to realize efficient phase locking.

\begin{figure*}[htbp]
    \centering
    \includegraphics[width=0.8\linewidth]{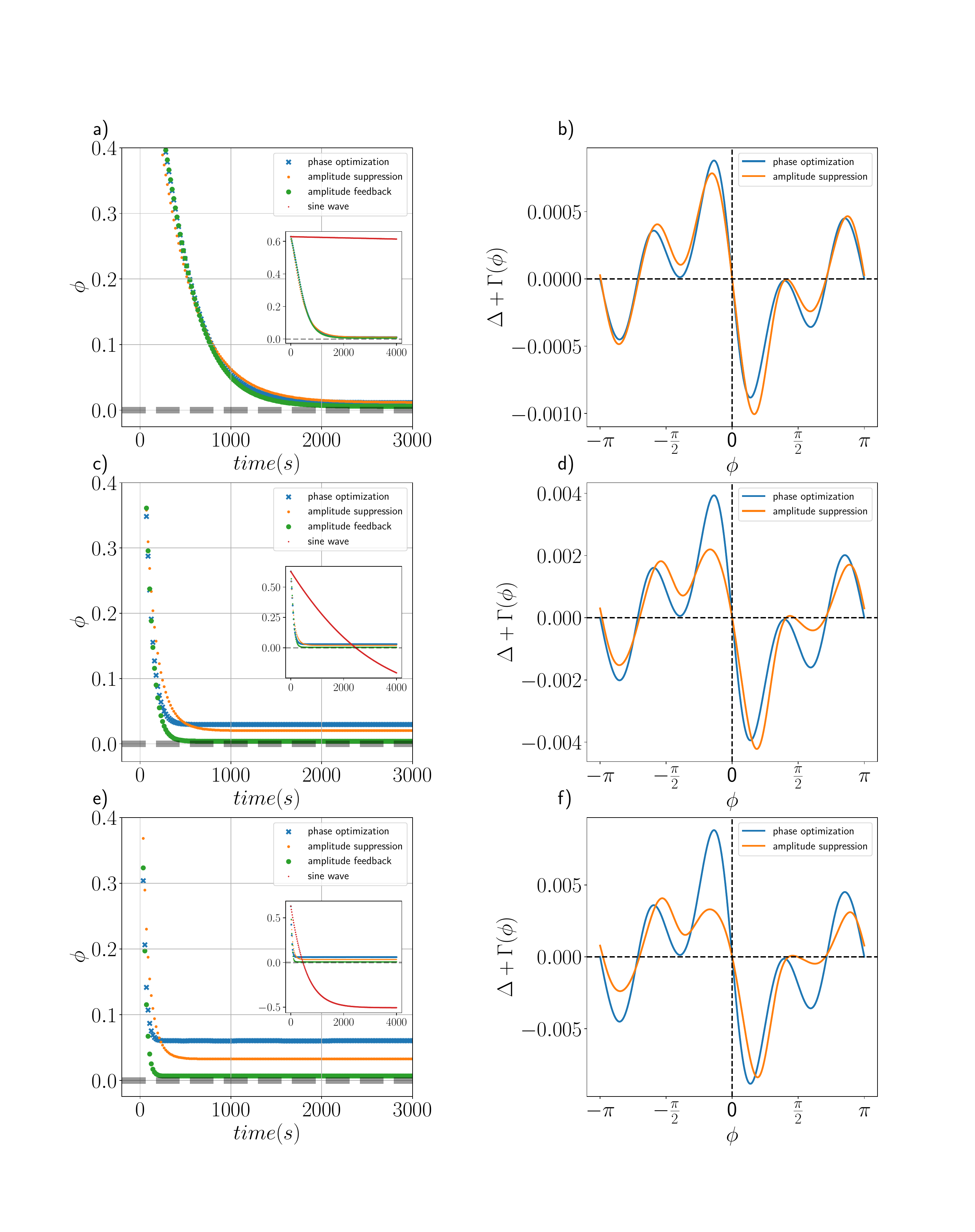}
	\caption{Phase locking with single-element control for three different input powers.
	  (a, c, e) Evolution of phase differences (calculated at every period). The insets show the results with sinusoidal inputs with equal powers for comparison.
	  (b, d, f) Corresponding phase coupling functions.
 	(a, b) $P=0.0005$, (c, d) $P=0.01$, and (e, f) $P=0.05$.
 	  }
    \label{fig:ring_one}
\end{figure*}

\begin{figure*}[htbp]
    \centering
    \includegraphics[width=0.8\linewidth]{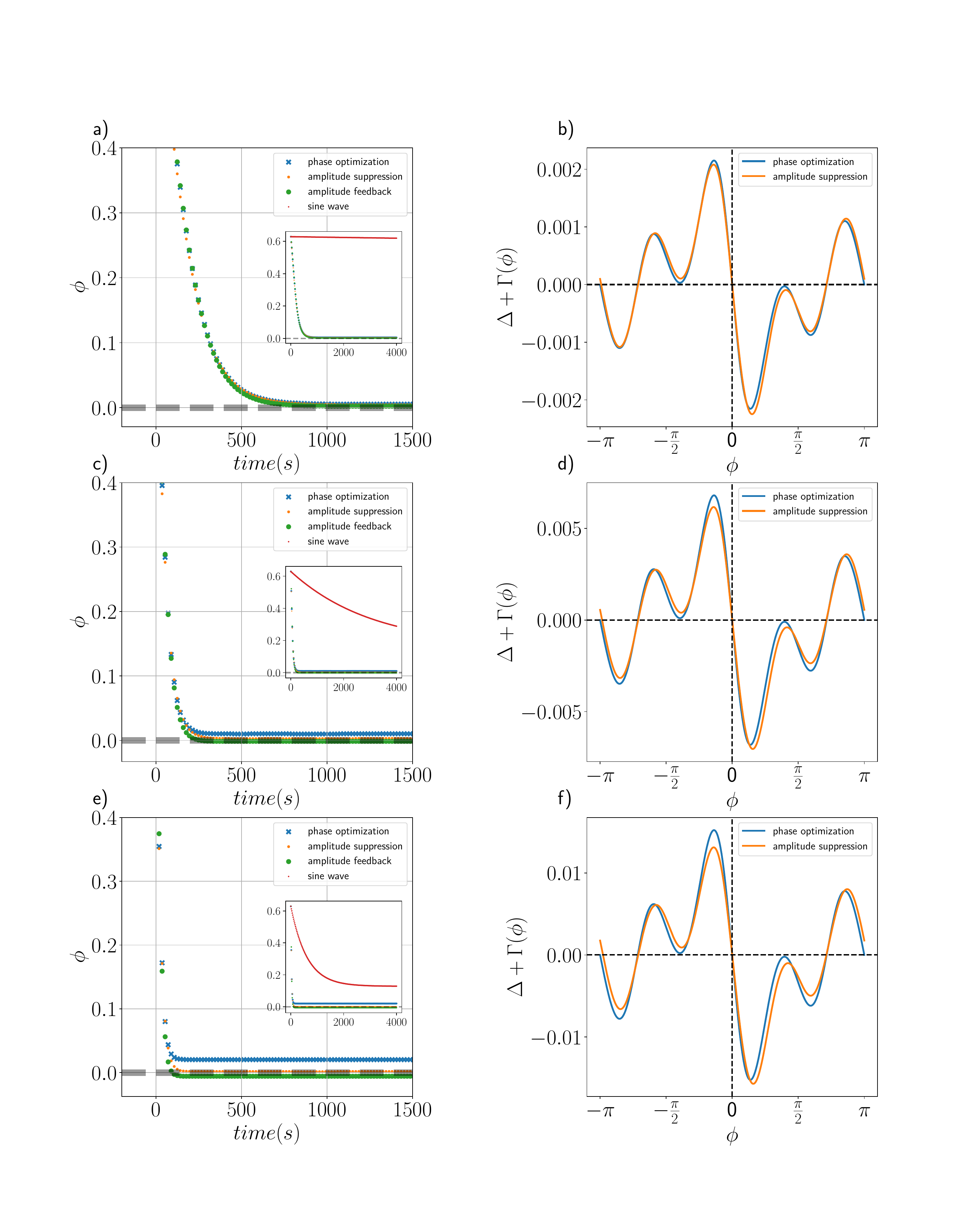}
	\caption{Phase locking with three-element control for three different input powers.
	  (a, c, e) Evolution of phase differences (calculated at every period).
	  The insets show the results with sinusoidal inputs with equal powers for comparison.
	  (b, d, f) Corresponding phase coupling functions.
 	(a, b) $P=0.001$, (c, d) $P=0.01$, and (e, f) $P=0.05$.
 	  }
    \label{fig:ring_three}
\end{figure*}

Figures~\ref{fig:ring_one} shows the results of optimal phase locking for the case (i) where only a single element ($i=1$) is controlled, and Fig.~\ref{fig:ring_three} shows the results for the case (ii) where three elements ($i=1, 2$ and $3$) of the network are simultaneously controlled.
We assume that there is no frequency mismatch, i.e., $\Delta=0$, and the target phase-locking point is $\phi^{*} = 0$. We use an amplitude penalty with the weight $W_{i}=30$ for all $i=1, ..., 10$ and set the gain for the amplitude feedback to $\alpha=30$.
The initial value of the relative phase is set as $\Delta \phi(0) = \pi / 5$ in order to realize phase locking at $\phi^*=0$.
The average input powers are set as $P = 0.0005$, $0.01$, and $0.05$ for the case (i), and $P = 0.001$, $0.01$ and $0.05$ for the case (ii). Note that $P$ is the net power of the inputs to the three elements in the case (ii).
In the insets of Figs.~\ref{fig:ring_one}(a, c, and e) and~\ref{fig:ring_three}(a, c, and e), the result for sinusoidal inputs of equivalent powers without amplitude suppression are shown as baselines in addition to the results with the  optimized inputs.

Figures~\ref{fig:ring_one}(b, d, and f) show the phase coupling functions for the optimized inputs with and without amplitude penalty for the case (i), where the functions $\Gamma(\phi)$ for varying values of the input power are plotted. Similarly, Figs.~\ref{fig:ring_three}(b, d, and f) show the phase coupling functions for the case (ii).
We can observe that the linear stability of the phase-locking point $\phi^*=0$ is lower in the case with amplitude penalty than in the case without amplitude penalty.
It is interesting to note that deterioration in the linear stability is large for the case (i), while it is much smaller for the case (ii).
Thus, the three-element control can more efficiently suppress the amplitude than the single-element control without lowering the linear stability.

Figures~\ref{fig:ring_one}(a, c, and e) show direct numerical simulations of the phase locking for the case (i) with single-element control.
We observe that for the weakest input $P=0.0005$, the relative phase converges to the target phase-locking point $\phi^{*}=0$ for all three methods successfully but slowly (though much faster than the simple sinusoidal input).
For intermediate and stronger inputs, $P=0.01$ and $P=0.05$, the phase-only method fails to converge to the target phase-locking point because the amplitude deviations from the limit cycle become too large and the validity of the reduced phase equation is not preserved. The amplitude penalty method is slightly better than the phase-only method, but it also fails to realize the target phase-locking point.
Only the amplitude feedback method successfully suppresses the amplitude deviations and realizes correct phase-locking point.

Finally, Figs.~\ref{fig:ring_three}(a, c, and e) show the results of simulations for the case (ii) with three-element control.
We observe that all three methods lead to convergence to the target phase-locking point when the average power of the input is the weakest, $P = 0.001$.
At the intermediate input, $P=0.01$, the phase-only method starts to deviate from the target, while the amplitude penalty method and amplitude feedback method correctly converge to the target.
At the largest input, $P=0.05$, the phase-only method shows considerable deviation from the target, while the amplitude penalty and feedback methods both successfully realize convergence to the target.

The improvement in the convergence speed in both cases with amplitude suppression is remarkable.
It is also interesting that the amplitude penalty method works much better in the present three-element control case (ii) than in the previous single-element control case (i). This is because there is greater flexibility in optimizing the input when three elements with different PSFs and ISFs can be used.

\subsection{Phase-locking of random network}

\begin{figure*}[htbp]
    \includegraphics[width=\linewidth]{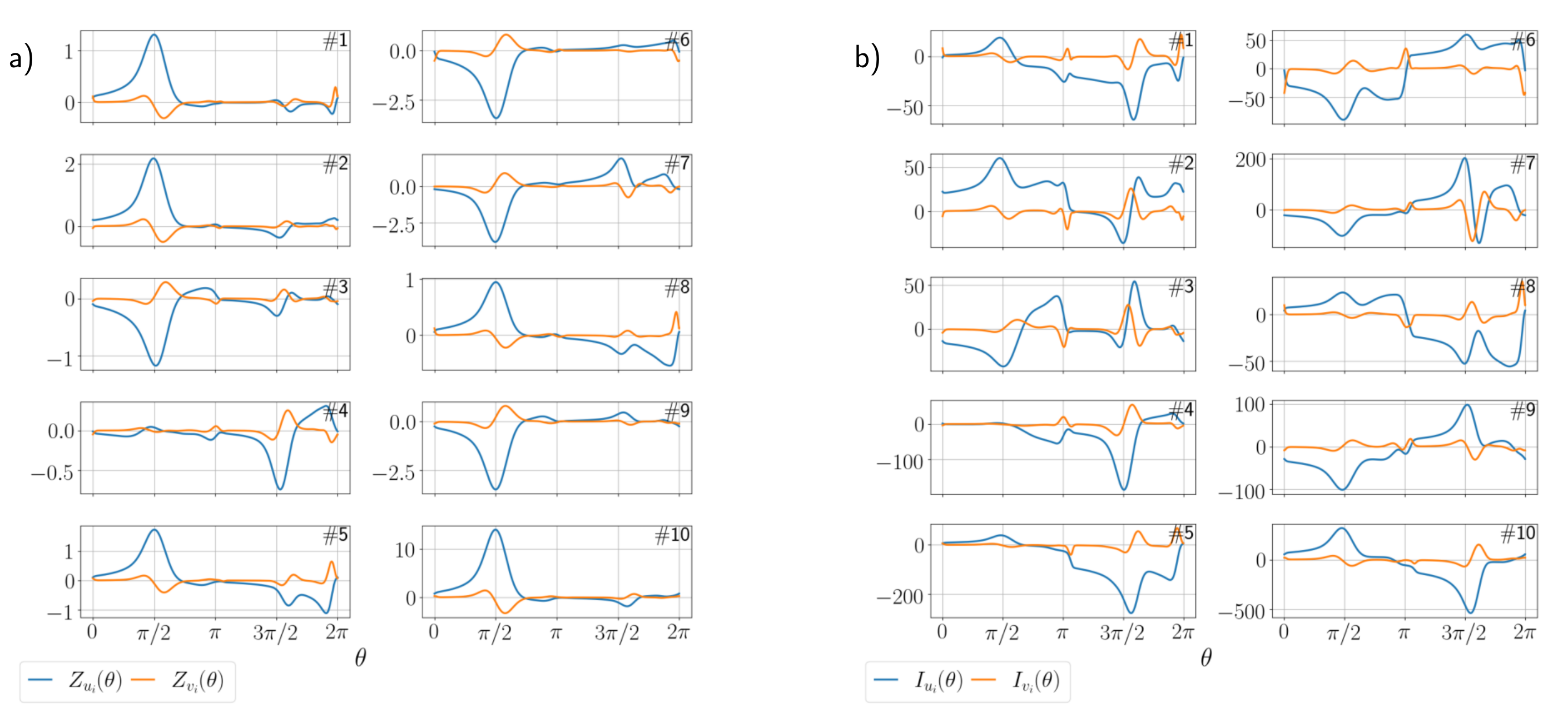}
	\caption{Phase and amplitude sensitivity functions of the random network.
	  (a) PSFs ${\bm z}_{i}(\theta) = (z_{i}^{u}, z_{i}^{v})$.
	  (b) ISFs ${\bm I}_{i}(\theta) = ({\bm I}_{i, u}, {\bm I}_{i, v})$.}
	\label{fig:random_psf_isf}
\end{figure*}

Numerically obtained PSFs $\{ {\bm z}_i\} $ and ISFs $\{ {\bm I}_i \}$ of the limit-cycle solution of the random network are shown in Figs.~\ref{fig:random_psf_isf}(a) and (b), respectively.
The PSFs of the networks are analyzed previously in~\cite{nakao2018phase}, while ISFs are newly analyzed in this study.

\begin{figure*}[htbp]
  \centering
    \includegraphics[width=0.75\linewidth]{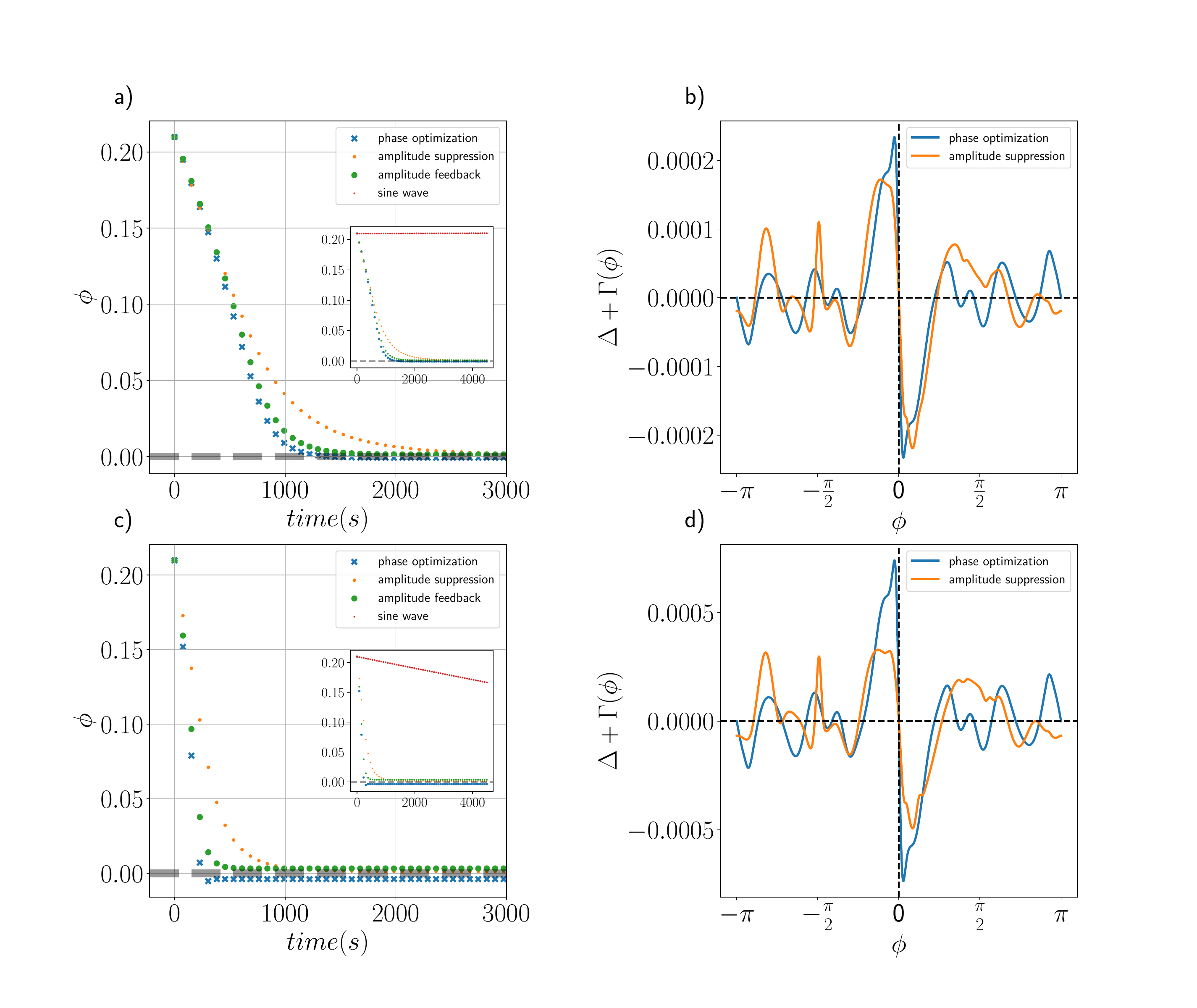}
	\caption{Phase locking with single-element control for two different input powers.
	  The insets show the results with sinusoidal inputs with equal powers for comparison.
	(a, c) Evolution of the phase differences (calculated at every period).
	(b, d) Corresponding phase coupling functions.
	(a,b) $P=5\cdot10^{-5}$, (c,d) $P=5 \cdot 10^{-4}$.}
    \label{fig:random_one}
\end{figure*}

\begin{figure*}[htbp]
  \centering
    \includegraphics[width=0.75\linewidth]{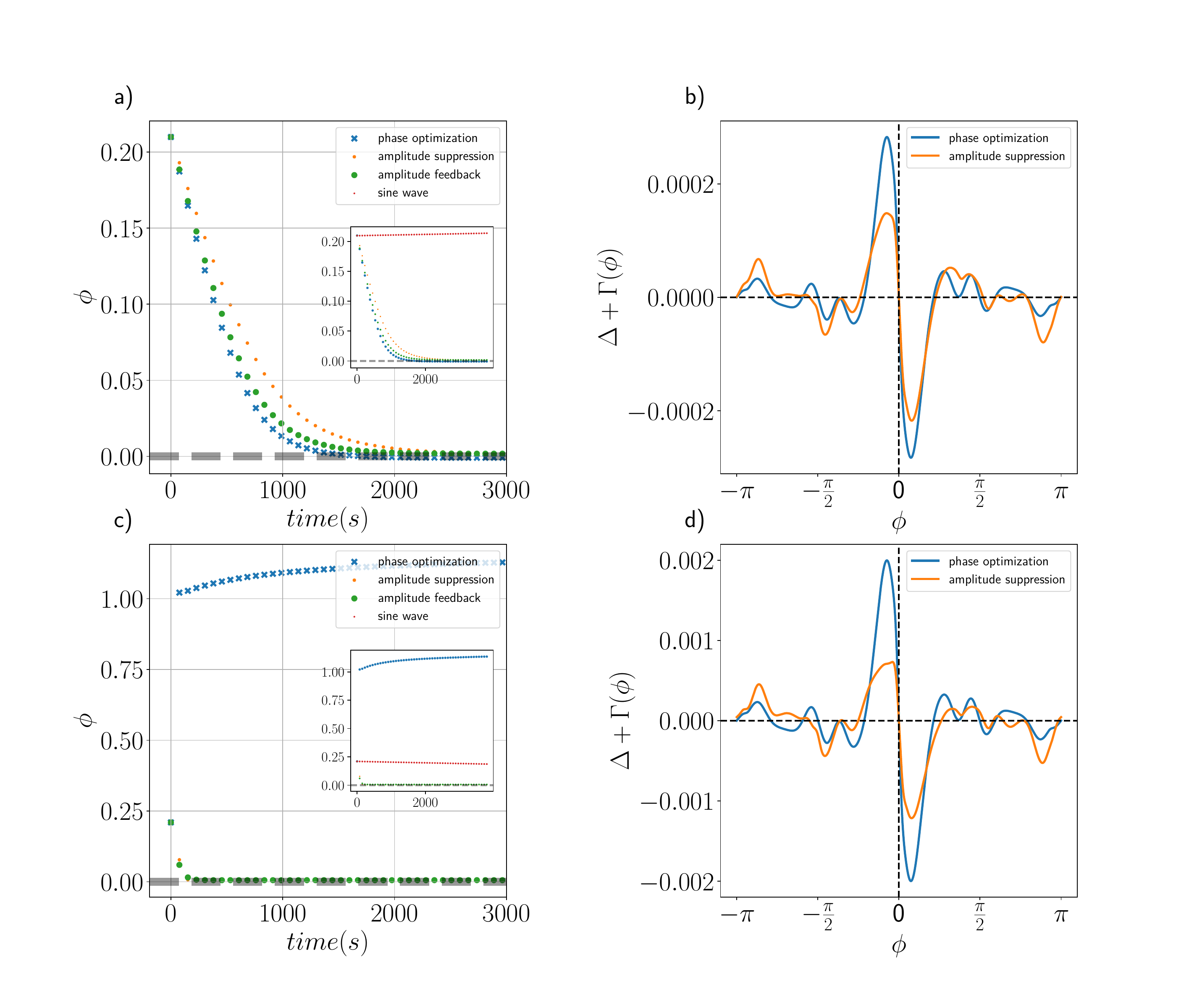}
	\caption{Phase locking with three-element control for two different input powers.
	(a, c) Evolution of the phase differences (calculated at every period).
	  The insets show the results with sinusoidal inputs with equal powers for comparison.
	(b, d) Corresponding phase coupling functions.
	(a,b) $P=1\cdot10^{-5}$, (c,d) $P=5 \cdot 10^{-4}$.}
    \label{fig:random_three}
\end{figure*}

Figures~\ref{fig:random_one} shows the results of optimal phase locking for the case (i) where only a single element ($i=1$) is controlled, and Fig.~\ref{fig:random_three} shows the results for the case (ii) where three elements ($i=1, 2$ and $3$) are simultaneously controlled.
In the insets of Figs.~\ref{fig:random_one}(a, c) and~\ref{fig:random_three}(a, c), the result for sinusoidal inputs of equivalent powers without amplitude suppression are shown as baselines in addition to the results with the  optimized inputs.
We again assume that there is no frequency mismatch $\Delta=0$ and set the target phase-locking point as $\phi^{*}= 0$. We set the weight of the amplitude penalty as $W_{i}=20$ and use a feedback gain of $\alpha = 500$. The initial value of the relative phase is set as $\Delta \phi(0) = \pi / 15$ in order to realize phase locking at $\phi^*=0$.
For the case (i) with single-element control, the power of the input is set as $P=5 \cdot 10^{-5}$ or $P=5 \cdot 10^{-4}$,
and for the case (ii) with three-element control, the net power of the inputs is set as $P=1 \cdot 10^{-5}$ or $P=5 \cdot 10^{-4} $.

Figures~\ref{fig:random_one}(b,d) show the phase coupling functions with and without amplitude penalty for the case (i), and similarly Figs.~\ref{fig:random_three}(b,d) show the phase coupling functions for the case (ii).
Due to the complex functional shapes of the PSFs, many stable phase-locking points coexist. We focus only on the target point $\phi^*=0$ in this study.
We can observe that the linear stability of $\phi^*=0$ is lower in the case with amplitude penalty than in the case without amplitude penalty.

Figures~\ref{fig:random_one}(a,c) show the results of direct numerical simulations for the case (i) with the single-element control. We observe that for the weaker input $P=5 \cdot 10^{-5}$, the relative phase converges to the target phase-locking point $\phi^{*}=0$ for all three methods correctly but slowly (though much faster than the simple sinusoidal input).
For the stronger input $P=5 \cdot 10^{-4}$, the phase-only method yields some deviations from the target point, while the amplitude penalty and amplitude feedback methods successfully realize the target phase-locking point.

Finally, Figs.~\ref{fig:random_three}(a,c) show the results of direct numerical simulations for the case (ii) with the three-element control.
All three methods lead to successful but slow convergence to the target phase-locking point when the input power is sufficiently weak, $P = 1 \cdot 10^{-5}$.
For the stronger input $P=5 \cdot 10^{-4}$, the phase-only method largely deviates from the target phase-locking point and converges to a far distant point within one period of the oscillation.
In contrast, both the amplitude penalty and amplitude feedback methods realize remarkably fast convergence to the correct target.
This example clearly demonstrates what could happen when the amplitude is not suppressed and the validity of the reduced phase equations is not preserved.

\section{Conclusion}

A general framework for the phase-amplitude reduction of collective oscillations in networked dynamical systems is formulated and applied to networks of coupled FitzHugh-Nagumo elements with two different topologies. It was shown that the reduced phase-amplitude equations can be used to derive optimal input waveforms for fast phase locking of the collective oscillations.
We observed that simple sinusoidal inputs yielded significantly slower convergence or even failure of convergence to the target phase-locking point, while the optimized inputs yielded improved convergence.
By further suppressing the amplitude deviations using the amplitude penalty or amplitude feedback method, we could realize even faster convergence to the target phase-locking point.
Moreover, it was shown that collective oscillations can be more efficiently controlled by applying inputs to multiple elements rather than to a single element.
These results elucidate the necessity and utility of optimized control schemes for efficient phase locking of collectively oscillating networks.
It would also be possible to extend the present method to such optimization problems as choosing the most suitable combination of elements to achieve stable phase locking or finding the combination of elements to avoid unnecessary phase-locking points other than those of interest.

\section*{Acknowledgement}
We thank Y. Kato and S. Takata for useful discussions. P. M. and H. N. acknowledges JSPS KAKENHI JP22K11919, JP22H00516, and JST CREST JP-MJCR1913 for financial support.
J. Z. acknowledges the National Natural Science Foundation of China (Grant No. 12202195) for financial support.

\appendix

\section{Floquet theory and adjoint equations}

In the main text, we derived coupled adjoint equations for the PSF and ISFs of the network elements. As discussed in Refs.~\cite{kuramoto2019concept,takata2021fast}, the PSF and ISFs of the limit cycles are the left Floquet eigenvectors of the oscillator. We here briefly explain their relations in the case of networked dynamical systems.

We use the collective expression $\dot{\bm X} = {\bm F}({\bm X})$ to describe the whole network, where ${\bm X} = ({\bm x}_1, ..., {\bm x}_N) \in {\mathbb R}^M$ and ${\bm F} \in {\mathbb R}^M$ is given by
\begin{align}
{\bm F}({\bm X}) = \left( {\bm f}_1({\bm x}_1) + \sum_{j=1}^N {\bm g}_{1j}({\bm x}_j, {\bm x}_1),
\ ...,\
{\bm f}_N({\bm x}_N) + \sum_{j=1}^N {\bm g}_{Nj}({\bm x}_j, {\bm x}_N)
\right).
\end{align}
We consider a network state ${\bm X}(t) = \tilde{\bm X}(t) + {\bm Y}(t)$ near the limit-cycle solution $\tilde{\bm X}(t) = (\tilde{\bm x}_1(t), ..., \tilde{\bm x}_N(t))$, where ${\bm Y}(t) = ({\bm y}_1(t), ..., {\bm y}_N(t))$ is a small variation (${\bm y}_i \in {\mathbb R}^{n_i}$).
The linearized dynamics of  the small variation ${\bm Y}(t)$ is expressed as
\begin{equation}
	\dot{\bm Y}(t) = J(t) {\bm Y}(t),
\end{equation}
where $J (t) = D {\bm F}(\tilde{\bm X}(t)) \in \mathbb R^{M \times M}$ is a $T$-periodic Jacobian matrix of the system's vector field ${\bm F}({\bm X})$ evaluated on the limit cycle at ${\bm X} = \tilde{\bm X}(t)$.
This $J(t)$ can be expressed as
\begin{align}
J(t) =
\begin{pmatrix}
J_{1}(t)+J_{11}(t) & J_{12}(t) & ... & J_{1N}(t)
\\
J_{21}(t) & J_2(t) + J_{22}(t) & ... & J_{2N}(t)
\\
\vdots & \vdots & \ddots & \vdots
\\
J_{N1}(t) & J_{N2}(t) & ... & J_N(t) + J_{NN}(t)
\end{pmatrix},
\end{align}
where $J_i(t)$ and $J_{ij}(t)$ are defined in the main text.

We denote the fundamental matrix of this linearized system as $\Psi(t) \in \mathbb R^{M \times M}$,
 which satisfies $\dot{\Psi}(t) = J(t) \Psi(t)$, with an initial condition $\Psi(0) = E$, where $E$ is the  $M\times M$ identity matrix.
According to Floquet theory, the fundamental matrix can be further represented as~\cite{guckenheimer2013nonlinear}
\begin{equation}
	\Psi(t) = Q(t) \exp(\Lambda t),
\end{equation}
where  $Q(t) \in {\mathbb R}^{M \times M}$ is a $T$-periodic matrix and $\Lambda \in {\mathbb R}^{M \times M}$ is a constant matrix.
 The eigenvalue problem for the matrix $\Lambda$ is given by
\begin{align}
	 & \Lambda {\bm U}_{m} = \lambda_{m}{\bm U}_{m},        \\
	 & \Lambda^{\dag}{\bm V}_{m} = \lambda^{*}_m {\bm V}_{m},
\end{align}
 for $m=0, ..., M-1$, where $\lambda_{m}$ is the $m$th eigenvalue (Floquet exponent), ${\bm U}_m = ( {\bm u}_1, ..., {\bm u}_N ) \in {\mathbb C}^M$ is the right eigenvector (${\bm u}_i \in {\mathbb R}^{n_i}$), ${\bm V}_m = ( {\bm v}_1, ..., {\bm v}_N ) \in {\mathbb C}^M$ is the left eigenvector (${\bm v}_i \in {\mathbb R}^{n_i}$), $\dag$ indicates transpose, and $*$ indicates complex conjugate.
The left and right eigenvectors ${\bm U}_m$ and ${\bm V}_m$ can be bi-orthogonalized to satisfy $\langle {\bm U}_m, {\bm V}_n \rangle = \delta_{mn}$, where $\delta$ is the Kronecker delta and $\langle {\bm A}, {\bm B} \rangle = \sum_{i=1}^M A_i^* B_i$ is an inner product of two vectors.

Note that these eigenvectors are also the left and right eigenvectors of $\Psi(T) = e^{\Lambda T}$, called the monodromy matrix, associated with the eigenvalues $e^{\lambda_m^* T}$ and $e^{\lambda_m T}$, respectively. Since ${\bm Y}(t) = \Psi(t) {\bm Y}(0)$, ${\bm Y}(T) = e^{\Lambda T} {\bm Y}(0)$. Thus, the eigenvalues characterize the linear stability of the limit cycle.
The Floquet exponents $\lambda_m$, are sorted in decreasing order of their real parts i.e., $ \lambda_{0} = 0 > Re(\lambda_{1} ) \geq Re(\lambda_{2}) \geq \ldots \geq Re(\lambda_{N-1})$, which are all negative except $\lambda_0$ because of the assumption that the limit cycle is exponentially stable.
The tangent vector $ (d/dt)\tilde{\bm X}(t)|_{t=0}$ of the limit cycle at $\tilde{\bm X}(0)$ is proportional to a right eigenvector of $\Psi(T)$ with $\lambda_0 = 0$, since it should remain the same after a one-period evolution.
In this study, we define ${\bm U}_0 = (1/\omega) (d/dt)\tilde{\bm X}(t)|_{t=0}$, where the scaling factor $1/\omega$ is introduced to be consistent with the definition of the asymptotic phase.

From the left and right eigenvectors  ${\bm U}_m$ and ${\bm V}_m$, we can derive a set of bi-orthogonal   $T$-periodic basis vectors $\bm U_m(t)  = Q(t) {\bm U}_m(0)$ and $\bm V_m(t)   = Q(t)^{-\dag} {\bm V}_m(0)$  for $0 \leq t \leq T$ satisfying the bi-orthogonality and periodicity condition i.e, $\langle \bm U_m(t), \bm V_{n}(t) \rangle = \delta_{mn}$.
We call these vectors ${\bm U}(t)$ and ${\bm V}(t)$ accordingly the right Floquet eigenvectors and left Floquet eigenvectors, respectively.

As shown in~\cite{takata2021fast}, the left and right Floquet eigenvectors satisfy the adjoint equations
\begin{align}
	 & \frac{d{\bm U}_m(t)}{dt} = [J(t) - \lambda_m] {\bm U}_m(t),            \\
	 & \frac{d{\bm V}_m(t)}{dt} = -[J^{\dag}(t) - \lambda^{*}_m] {\bm V}_m(t),
\end{align}
with initial conditions
\begin{equation}
	{\bm U}_m(0) = {\bm U}_m \quad {\bm V}_m(0) = {\bm V}_m.
\end{equation}
We note that the adjoint equation for the left eigenvector ${\bm V}_m(t)$ gives the set of adjoint equations for ${\bm z}_i$ ($m=0$) and ${\bm I}_{m,i}$ ($m=1, ..., M-1$) in the main text.
The PSF and ISFs are given by the left Floquet eigenvectors as
\begin{align}
{\bm Z}(\theta) = {\bm V}_0(\theta / \omega),
\quad
{\bm I}_m(\theta) = {\bm V}_m(\theta / \omega)^*,
\end{align}
for $m=1, ..., M-1$.

\section{Coupling matrix for the random network}

The coupling matrix for the random network is given by
\begin{align}
K=
\begin{pmatrix}
0.000 & 0.409 & -0.176 & -0.064 & -0.218 & 0.464 & -0.581 & 0.101 & -0.409 & -0.140 \\
0.229 & 0.000 & 0.480 & -0.404 & -0.409 & 0.040 & 0.125 & 0.099 & -0.276 & -0.131 \\
-0.248 & 0.291 & 0.000 & -0.509 & -0.114 & 0.429 & 0.530 & 0.195 & 0.416 & -0.597 \\
-0.045 & 0.039 & 0.345 & 0.000 & 0.579 & -0.232 & 0.121 & 0.130 & -0.345 & 0.463 \\
-0.234 & -0.418 & -0.195 & -0.135 & 0.000 & 0.304 & 0.124 & 0.038 & -0.049 & 0.183 \\
-0.207 & 0.536 & -0.158 & 0.533 & -0.591 & 0.000 & -0.273 & -0.571 & 0.110 & -0.354 \\
0.453 & -0.529 & -0.287 & -0.237 & 0.470 & -0.002 & 0.000 & -0.256 & 0.438 & 0.211 \\
-0.050 & 0.552 & 0.330 & -0.148 & -0.326 & -0.175 & -0.240 & 0.000 & 0.263 & 0.079 \\
0.389 & -0.131 & 0.383 & 0.413 & -0.383 & 0.532 & -0.090 & 0.025 & 0.000 & 0.496 \\
0.459 & 0.314 & -0.121 & 0.226 & 0.314 & -0.114 & -0.450 & -0.018 & -0.333 & 0.000 \\
\end{pmatrix}.
\label{mat:random_network_adjacency}
\end{align}

\bibliographystyle{unsrt}
\bibliography{refs}

\begin{thebibliography}{10}

\bibitem{winfree1967biological}
Arthur~T Winfree.
\newblock Biological rhythms and the behavior of populations of coupled
  oscillators.
\newblock {\em Journal of theoretical biology}, 16(1):15--42, 1967.

\bibitem{winfree1980geometry}
Arthur~T Winfree.
\newblock {\em The geometry of biological time}, volume~2.
\newblock Springer, 1980.

\bibitem{pikovsky2002synchronization}
Arkady Pikovsky, Michael Rosenblum, and J{\"u}rgen Kurths.
\newblock Synchronization: a universal concept in nonlinear science, 2002.

\bibitem{strogatz2012sync}
Steven~H Strogatz.
\newblock {\em Sync: How order emerges from chaos in the universe, nature, and
  daily life}.
\newblock Hachette UK, 2012.

\bibitem{kuramoto1984chemical}
Yoshiki Kuramoto.
\newblock Chemical turbulence.
\newblock In {\em Chemical oscillations, waves, and turbulence}, pages
  111--140. Springer, 1984.

\bibitem{hoppensteadt1997weakly}
Frank~C Hoppensteadt and Eugene~M Izhikevich.
\newblock {\em Weakly connected neural networks}, volume 126.
\newblock Springer Science \& Business Media, 1997.

\bibitem{ermentrout2010mathematical}
Bard Ermentrout and David~H Terman.
\newblock {\em Mathematical foundations of neuroscience}, volume~35.
\newblock Springer, 2010.

\bibitem{brown2004phase}
Eric Brown, Jeff Moehlis, and Philip Holmes.
\newblock On the phase reduction and response dynamics of neural oscillator
  populations.
\newblock {\em Neural computation}, 16(4):673--715, 2004.

\bibitem{nakao2016phase}
Hiroya Nakao.
\newblock Phase reduction approach to synchronisation of nonlinear oscillators.
\newblock {\em Contemporary Physics}, 57(2):188--214, 2016.

\bibitem{monga2019phase}
Bharat Monga, Dan Wilson, Tim Matchen, and Jeff Moehlis.
\newblock Phase reduction and phase-based optimal control for biological
  systems: a tutorial.
\newblock {\em Biological cybernetics}, 113(1-2):11--46, 2019.

\bibitem{monga2019optimal}
Bharat Monga and Jeff Moehlis.
\newblock Optimal phase control of biological oscillators using augmented phase
  reduction.
\newblock {\em Biological cybernetics}, 113(1-2):161--178, 2019.

\bibitem{iatsenko2013evolution}
Dima Iatsenko, Alan Bernjak, Tomislav Stankovski, Yuri Shiogai, P~Jane
  Owen-Lynch, PBM Clarkson, Peter~VE McClintock, and Aneta Stefanovska.
\newblock Evolution of cardiorespiratory interactions with age.
\newblock {\em Philosophical Transactions of the Royal Society A: Mathematical,
  Physical and Engineering Sciences}, 371(1997):20110622, 2013.

\bibitem{moehlis2006optimal}
Jeff Moehlis, Eric Shea-Brown, and Herschel Rabitz.
\newblock Optimal inputs for phase models of spiking neurons.
\newblock {\em Journal of computational and nonlinear dynamics}, 1(4):358--367,
  2006.

\bibitem{tanaka2014optimal}
Hisa-Aki Tanaka.
\newblock Optimal entrainment with smooth, pulse, and square signals in weakly
  forced nonlinear oscillators.
\newblock {\em Physica D: Nonlinear Phenomena}, 288:1--22, 2014.

\bibitem{tanaka2015optimal}
Hisa-Aki Tanaka, Isao Nishikawa, J{\"u}rgen Kurths, Yifei Chen, and
  Istv{\'a}n~Z Kiss.
\newblock Optimal synchronization of oscillatory chemical reactions with
  complex pulse, square, and smooth waveforms signals maximizes tsallis
  entropy.
\newblock {\em Europhysics Letters}, 111(5):50007, 2015.

\bibitem{harada2010optimal}
Takahiro Harada, Hisa-Aki Tanaka, Michael~J Hankins, and Istv{\'a}n~Z Kiss.
\newblock Optimal waveform for the entrainment of a weakly forced oscillator.
\newblock {\em Physical review letters}, 105(8):088301, 2010.

\bibitem{kato2021optimization}
Yuzuru Kato, Anatoly Zlotnik, Jr-Shin Li, and Hiroya Nakao.
\newblock Optimization of periodic input waveforms for global entrainment of
  weakly forced limit-cycle oscillators.
\newblock {\em Nonlinear dynamics}, 105(3):2247--2263, 2021.

\bibitem{zlotnik2013optimal}
Anatoly Zlotnik, Yifei Chen, Istv{\'a}n~Z Kiss, Hisa-Aki Tanaka, and Jr-Shin
  Li.
\newblock Optimal waveform for fast entrainment of weakly forced nonlinear
  oscillators.
\newblock {\em Physical review letters}, 111(2):024102, 2013.

\bibitem{watanabe2019optimization}
Nobuhiro Watanabe, Yuzuru Kato, Sho Shirasaka, and Hiroya Nakao.
\newblock Optimization of linear and nonlinear interaction schemes for stable
  synchronization of weakly coupled limit-cycle oscillators.
\newblock {\em Physical Review E}, 100(4):042205, 2019.

\bibitem{pikovsky2015maximizing}
Arkady Pikovsky.
\newblock Maximizing coherence of oscillations by external locking.
\newblock {\em Physical Review Letters}, 115(7):070602, 2015.

\bibitem{zlotnik2016phase}
Anatoly Zlotnik, Raphael Nagao, Istv{\'a}n~Z Kiss, and Jr-Shin Li.
\newblock Phase-selective entrainment of nonlinear oscillator ensembles.
\newblock {\em Nature communications}, 7(1):10788, 2016.

\bibitem{wedgwood2013phase}
Kyle~CA Wedgwood, Kevin~K Lin, Ruediger Thul, and Stephen Coombes.
\newblock Phase-amplitude descriptions of neural oscillator models.
\newblock {\em The Journal of Mathematical Neuroscience}, 3:1--22, 2013.

\bibitem{wilson2016isostable}
Dan Wilson and Jeff Moehlis.
\newblock Isostable reduction of periodic orbits.
\newblock {\em Physical Review E}, 94(5):052213, 2016.

\bibitem{shirasaka2017phase}
Sho Shirasaka, Wataru Kurebayashi, and Hiroya Nakao.
\newblock Phase-amplitude reduction of transient dynamics far from attractors
  for limit-cycling systems.
\newblock {\em Chaos: An Interdisciplinary Journal of Nonlinear Science},
  27(2):023119, 2017.

\bibitem{mauroy2018global}
Alexandre Mauroy and Igor Mezi{\'c}.
\newblock Global computation of phase-amplitude reduction for limit-cycle
  dynamics.
\newblock {\em Chaos: An Interdisciplinary Journal of Nonlinear Science},
  28(7):073108, 2018.

\bibitem{wilson2019phase}
Dan Wilson and Bard Ermentrout.
\newblock Phase models beyond weak coupling.
\newblock {\em Physical review letters}, 123(16):164101, 2019.

\bibitem{mauroy2020koopman}
Alexandre Mauroy, Y~Susuki, and I~Mezi{\'c}.
\newblock {\em Koopman operator in systems and control}.
\newblock Springer, 2020.

\bibitem{shirasaka2020phase}
Sho Shirasaka, Wataru Kurebayashi, and Hiroya Nakao.
\newblock Phase-amplitude reduction of limit cycling systems.
\newblock {\em The Koopman Operator in Systems and Control: Concepts,
  Methodologies, and Applications}, pages 383--417, 2020.

\bibitem{takata2021fast}
Shohei Takata, Yuzuru Kato, and Hiroya Nakao.
\newblock Fast optimal entrainment of limit-cycle oscillators by strong
  periodic inputs via phase-amplitude reduction and floquet theory.
\newblock {\em Chaos: An Interdisciplinary Journal of Nonlinear Science},
  31(9):093124, 2021.

\bibitem{wilson2020phase}
Dan Wilson.
\newblock Phase-amplitude reduction far beyond the weakly perturbed paradigm.
\newblock {\em Phys. Rev. E}, 101:022220, Feb 2020.

\bibitem{kotani2020nonlinear}
Kiyoshi Kotani, Yutaro Ogawa, Sho Shirasaka, Akihiko Akao, Yasuhiko Jimbo, and
  Hiroya Nakao.
\newblock Nonlinear phase-amplitude reduction of delay-induced oscillations.
\newblock {\em Physical Review Research}, 2(3):033106, 2020.

\bibitem{mauroy2013isostables}
Alexandre Mauroy, Igor Mezi{\'c}, and Jeff Moehlis.
\newblock Isostables, isochrons, and koopman spectrum for the action--angle
  representation of stable fixed point dynamics.
\newblock {\em Physica D: Nonlinear Phenomena}, 261:19--30, 2013.

\bibitem{mauroy2016global}
Alexandre Mauroy and Igor Mezi{\'c}.
\newblock Global stability analysis using the eigenfunctions of the koopman
  operator.
\newblock {\em IEEE Transactions on Automatic Control}, 61(11):3356--3369,
  2016.

\bibitem{kawamura2008collective}
Yoji Kawamura, Hiroya Nakao, Kensuke Arai, Hiroshi Kori, and Yoshiki Kuramoto.
\newblock Collective phase sensitivity.
\newblock {\em Physical review letters}, 101(2):024101, 2008.

\bibitem{kori2009collective}
Hiroshi Kori, Yoji Kawamura, Hiroya Nakao, Kensuke Arai, and Yoshiki Kuramoto.
\newblock Collective-phase description of coupled oscillators with general
  network structure.
\newblock {\em Physical Review E}, 80(3):036207, 2009.

\bibitem{nakao2014phase}
Hiroya Nakao, Tatsuo Yanagita, and Yoji Kawamura.
\newblock Phase-reduction approach to synchronization of spatiotemporal rhythms
  in reaction-diffusion systems.
\newblock {\em Physical review X}, 4(2):021032, 2014.

\bibitem{nakao2018phase}
Hiroya Nakao, Sho Yasui, Masashi Ota, Kensuke Arai, and Yoji Kawamura.
\newblock Phase reduction and synchronization of a network of coupled dynamical
  elements exhibiting collective oscillations.
\newblock {\em Chaos: An Interdisciplinary Journal of Nonlinear Science},
  28(4):045103, 2018.

\bibitem{nakao2021sparse}
Hiroya Nakao, Katsunori Yamaguchi, Shingo Katayama, and Tatsuo Yanagita.
\newblock Sparse optimization of mutual synchronization in collectively
  oscillating networks.
\newblock {\em Chaos: An Interdisciplinary Journal of Nonlinear Science},
  31(6):063113, 2021.

\bibitem{nakao2021phase}
Hiroya Nakao.
\newblock Phase and amplitude description of complex oscillatory patterns in
  reaction-diffusion systems.
\newblock In {\em Physics of Biological Oscillators: New Insights into
  Non-Equilibrium and Non-Autonomous Systems}, pages 11--27. Springer, 2021.

\bibitem{izhikevich2006polychronization}
Eugene~M Izhikevich.
\newblock Polychronization: computation with spikes.
\newblock {\em Neural computation}, 18(2):245--282, 2006.

\bibitem{kuramoto2019concept}
Yoshiki Kuramoto and Hiroya Nakao.
\newblock On the concept of dynamical reduction: the case of coupled
  oscillators.
\newblock {\em Philosophical Transactions of the Royal Society A},
  377(2160):20190041, 2019.

\bibitem{guckenheimer2013nonlinear}
John Guckenheimer and Philip Holmes.
\newblock {\em Nonlinear oscillations, dynamical systems, and bifurcations of
  vector fields}, volume~42.
\newblock Springer Science \& Business Media, 2013.

\bibitem{press2007numerical}
William~H Press.
\newblock {\em Numerical recipes 3rd edition: The art of scientific computing}.
\newblock Cambridge university press, 2007.

\bibitem{shirasaka2017optimizing}
Sho Shirasaka, Nobuhiro Watanabe, Yoji Kawamura, and Hiroya Nakao.
\newblock Optimizing stability of mutual synchronization between a pair of
  limit-cycle oscillators with weak cross coupling.
\newblock {\em Physical Review E}, 96(1):012223, 2017.

\bibitem{singhal2023optimal}
Bharat Singhal, Istv{\'a}n~Z Kiss, and Jr-Shin Li.
\newblock Optimal phase-selective entrainment of heterogeneous oscillator
  ensembles.
\newblock {\em SIAM Journal on Applied Dynamical Systems}, 22(3):2180--2205,
  2023.

\end{thebibliography}
\newpage

\end{document}